\documentclass[a4paper,fleqn,usenatbib,useAMS]{mnras}

\usepackage{graphicx}	
\usepackage{amsmath}	
\usepackage{amssymb}	
\usepackage{multicol}        
\usepackage{bm}		
\usepackage{pdflscape}	

\usepackage[T1]{fontenc}
\usepackage{ae,aecompl}
\usepackage{mathptmx}

\usepackage{hyperref}

\newcommand{\kms}{\,km\,s$^{-1}$} 

\title[Semi-analytic spectral fitting]{Semi-analytic spectral fitting: simultaneously modelling the mass accumulation and chemical evolution in MaNGA spiral galaxies}

\author[S. Zhou]{Shuang Zhou$^{1}$\thanks{Contact e-mail: \href{mailto:Shuang.Zhou@nottingham.ac.uk}{Shuang.Zhou@nottingham.ac.uk}},
Michael Merrifield,$^{1}$
Alfonso Arag{\'o}n-Salamanca,$^{1}$
\\
$^{1}$School of Physics \& Astronomy, University of Nottingham, University Park, Nottingham, NG7 2RD, UK\\
}
\date{Last updated ???; in original form ???}

\pubyear{2022}

\begin{document}

\label{firstpage}
\pagerange{\pageref{firstpage}--\pageref{lastpage}}
\maketitle

\begin{abstract}
We develop a novel semi-analytic spectral fitting approach to quantify the star-formation histories (SFHs) and chemical enrichment histories (ChEHs) of individual galaxies. We construct simple yet general chemical evolution models that account for gas inflow and outflow processes as well as star formation, to investigate the evolution of merger-free star-forming systems. These models are fitted directly to galaxies' absorption-line spectra, while their emission lines are used to constrain current gas phase metallicity and star formation rate. We apply this method to spiral galaxies selected from the SDSS-IV MaNGA survey. By fitting the co-added absorption-line spectra for each galaxy, and using the emission-line constraints on present-day metallicity and star formation, we reconstruct both the SFHs and the ChEHs for all objects in the sample.  We can use these reconstructions to obtain archaeological measures of derived correlations such as the mass--metallicity relation at any redshift, which compare favourably with direct observations. We find that both the SFHs and ChEHs have strong mass dependence: massive galaxies accumulate their stellar masses and become enriched earlier. This mass dependence causes the observed flattening of the mass--metallicity relation at lower redshifts. The model also reproduces the observed gas-to-stellar mass ratio and its mass dependence. Moreover, we are able to determine that more massive galaxies have earlier gas infall times and shorter infall time-scales, and that the early chemical enrichment of low-mass galaxies is suppressed by strong outflows, while outflows are not very significant in massive galaxies.
   
\end{abstract}



\section{Introduction}
The evolution of a galaxy involves complex physical processes including gas inflow and outflow, star formation and metallicity enrichment processes, stellar and AGN feedback, and even external processes such as mergers and interactions between galaxies. As footprints of this galaxy evolution, the star-formation history (SFH) and chemical enrichment history (ChEH) of a galaxy provide important evidence for how those mechanisms operate during its evolution. Properly determining those properties is thus of great significance in galaxy evolution studies and is now attracting increasing attention.
 
The accumulation histories of mass and metals in the stellar component of a galaxy are often tracked through modelling its spectral energy distribution (SED) with the stellar population synthesis (SPS) approach of fitting model spectra. 
Investigations done in this manner show that the SFH of a galaxy is tightly correlated with its total stellar mass, in that massive galaxies tend to form their stellar masses earlier than less massive ones \citep{Panter2003,Kauffmann2003,Heavens2004,Panter2007,Fontanot2009,Peng2010,Muzzin2013,Peterken2020}. This so-called `down-sizing' formation is also seen in the accumulation of metals. Massive galaxies are found to have generally higher average metallicities than low mass ones \citep[e.g.][]{Gallazzi2005,Panter2007,Thomas2010}. More recent approaches extend the investigation of galaxies to detail the distribution of chemical compositions \citep[e.g.][]{Greener2021} or the full ChEHs \citep[e.g.][]{CampsFarina2021CALIFA}. Many of those studies rely on decomposing the spectrum of a galaxy into linear combinations of simple stellar population (SSP) templates with different ages and metallicities \citep{STARLIGHT,Cappellari2004,Wilkinson2017}. The SFH and ChEH of a galaxy derived in this way avoid assuming any specific functional forms, and thus impose the fewest assumptions. The downside 
to this approach is that this decomposition suffers from severe degeneracies. Adopting some kind of regularization \citep{Cappellari2004} can mitigate the problem, but it may also complicate the interpretation of the results.
Alternative solutions can be found in forward-modelling approaches, which incorporate models of SFH and ChEH to calculate composite stellar populations (CSPs), and compare them with the observed spectra to constrain models \citep{Noll2009,Chevallard2016,Carnall2018,Johnson2021}.
Results derived in this way naturally provide the information regarding the physical processes that motivate the SFH and ChEH models, but this approach suffers from model dependence \citep[e.g.][]{Zhou2020}. Moreover, although a detailed modelling of the ChEH is in principle possible in forward modelling, in practice most of the studies to-date adopt a single metallicity for all of the stellar populations.

Complementing the inference based on stellar populations, information can also be extracted from the gaseous component. The instantaneous star formation rate (SFR) can be estimated by measuring the strengths of emission lines produced by the galaxy \citep{Kennicutt1998ARA}. In addition, the metal content of ionized gas, which can be traced by emission line ratios \citep[e.g.][]{Tremonti2004,PP04,Izotov2006,Maiolino2008,Sanchez2013}, offers a snapshot of the chemical composition of the material from which stars are currently forming. Mirroring the results from stellar populations, \cite{Lequeux1979} found that massive galaxies have higher gas phase metallicities in the local Universe, and subsequently similar results have been found in large-scale surveys \citep[e.g.][]{Tremonti2004,Kewley2008,Andrews2013}. More recent works \citep[e.g.][]{Ly2016,Gillman2021,Sanders2021,Topping2021} extend this mass--metallicity relationship (MZR) to higher redshifts, allowing its evolution to be studied. Compared to stellar populations inferred from absorption lines, emission-line properties are easier to measure, and 
avoid some of the uncertainties present in stellar population synthesis modelling \citep{Conroy2009-FSPS1}, though they have some drawbacks of their own \citep{Lopez-Sanchez2012,Maiolino2019}. In addition, the gas component can only provide direct information on the current state of the galaxy; the full evolutionary picture can only be inferred by observing samples of galaxies at different redshifts. A combination of constraints from the stellar and gas phases should provide a more complete picture of the full evolutionary history, but to-date they have largely been treated separately.

Efforts have also been made to understand the physics behind the observed ChEHs and SFHs using chemical evolution models that incorporate a range of physical processes. Their predictions can then be compared to what we observe in the Universe. For example, \cite{Yates2012} combined the semi-analytic model L-GALAXIES with a chemical evolution model to reproduce the observed MZR of galaxies' gas component, but interestingly their model predicted a stellar metallicity significantly higher than the observations. \cite{Spitoni2017} proposed analytical models and successfully reproduced the MZR of stars in local galaxies from SDSS. \cite{Lian2018mzr} combined the constraints from the MZR of both gas and stars, and found that a strong outflow or variable IMF is needed to reproduce both relations. While these models can explain in a statistical sense the MZR of samples of galaxies, they are not designed to explore the diversity of individual objects, nor are they often used to study how the SFHs of these systems correlate with their ChEHs, and what they imply for the underlying physical processes. 

In this work we present a novel "semi-analytic spectral fitting" approach to investigate the evolution of both the gas and stellar phases in galaxies simultaneously. In contrast to previous works, we seek to model not only the global MZR, but also the evolution of every individual galaxy. This is achieved by incorporating a chemical evolution model into the spectral analysis code  Bayesian Inference for Galaxy Spectra ({\tt BIGS}), which has previously been successfully used to constrain the SFHs \citep{Zhou2020} and IMFs \citep{Zhou2019} for different types of galaxies. Incorporating the chemical evolution model allows us to self-consistently model both the SFH and ChEH of a galaxy; the forward modelling method combined with a Bayesian inference approach allows us to constrain this model using the full observed stellar and gas emission from the galaxy. The model we have adopted has been chosen to be reasonably simple in order to be computationally tractable and not have too many free parameters, but also to capture the major physical processes that are believed to be important in driving galaxies' chemical evolution.  

Having developed this new method, we apply it to spiral galaxies selected from the Mapping Nearby Galaxies at Apache Point Observatory (MaNGA; \citealt{Bundy2015}) survey. MaNGA provides a large number of spectra from across the faces of individual galaxies, which can be stacked to give a global spectrum of each object, providing the high signal-to-noise ratio (SNR) that is essential for determining the detailed ChEH and SFH of the whole galaxy. In addition, MaNGA's large sample size (over 10,000 galaxies) allows us to select a large enough sub-sample of spiral galaxies to explore how the ChEH and SFH, and the physical processes behind them, might vary between objects.

The paper is organised as follows. The method, including the chemical evolution model and fitting approach, is presented in \S\ref{sec:method}. We have tested this method extensively, and some of the key tests are shown in \S\ref{sec:test}. In \S\ref{sec:data} we introduce data sample from MaNGA. The main results of the paper are shown and discussed in \S\ref{sec:results}, and the interpretation of those results is given in \S\ref{sec:interp}. Our conclusions are summarized in \S\ref{sec:summary}. Throughout this work we use a standard $\Lambda$CDM cosmology with $\Omega_{\Lambda}=0.7$, 
$\Omega_{\rm M}=0.3$ and $H_0$=70\kms Mpc$^{-1}$.

\section{Method}
\label{sec:method}

The semi-analytic spectral fitting approach comprises two major ingredients:
\begin{itemize}
    \item a chemical evolution model that describes the physical processes driving the evolution of the galaxy;
    \item a dedicated tool that converts the evolution state of the galaxy into observables, such as the integrated spectrum, gas metallicity and star formation rate, and compares those model predictions to observed data so as to fit model parameters.    
\end{itemize}
In modelling the evolution of the gas, we seek a compromise that is sufficiently simple to be analytically tractable and mininimizes the number of free parameters, while being complex enough to capture the essential physical processes of gas inflow and outflow as well as varying star-formation activity. This generalized parametric model is combined with a stellar population synthesis approach to predict the galaxy spectrum, which is then compared with observations in a Bayesian context. In this section, we discuss these two ingredients in detail.

\subsection{The chemical evolution model}
The chemical evolution of a galaxy is dominated by complex baryonic processes including gas accretion from external sources, star formation and return of the chemical-enriched gas, and gas outlow due to feedback. Models characterising the chemical evolution of galaxies through analytical functions have been widely explored over the past 40 years. The simplest is the so-called "closed box" model, which considered the evolution of galaxies formed in an isolated gas cloud \citep[e.g.][]{Talbot1971,Tinsley1974}. With the rapid development of both modelling and observations, it was soon realized that the evolution of many galaxies deviates from such a simple picture, and gas flows need to be incorporated in the models \cite[e.g.][]{Tinsley1980,Chiosi1980,Lacey1985,Pagel1997,Erb2008,Recchi2008}. Following this historical development, we first attempted to model the evolution of our data with a closed box description, and test whether such a model can fit the many constraints in real observations. As the simple model fails to reproduce the complex behaviour of galaxies, inflows and outflows, as well as their time dependence, are sequentially incorporated into the model to increase its ability to match the observations. Finally, we end up with a simple yet flexible model, which we test widely to assess its power in capturing the main physics during the formation of a wide variety of galaxies. The major ingredients of the model and how its behaviour is driven by different parameters are discussed in what follows.

\subsubsection{Model ingredients}
\label{sec:ingredients}

Mathematically, the evolution of gas mass in a galaxy can be described by the following equation:
\begin{equation}
\dot{M}_{\rm g}(t)=\dot{M}_{\rm in}(t)-\psi(t)+\dot{M}_{\rm re}(t)-\dot{M}_{\rm out}(t).
\end{equation}
Here, $\dot{M}_{\rm in}(t)$ and $\dot{M}_{\rm out}(t)$ are the gas inflows and outflows respectively, while $\psi(t)$ and $\dot{M}_{\rm re}(t)$ characterise the star formation processes and the mass ejection from dying stars. There is a wealth of poorly-understood physics behind these functions, so here we adopt simple parametric models for them that seek to capture its essence. For the gas infall, we simply assume an exponentially decaying rate
\begin{equation}
\dot{M}_{\rm in}(t)=A e^{-(t-t_0)/\tau}, \ \ \ t > t_0
\end{equation}
where $t_0$ is the time that gas begins to infall, $\tau$ parameterizes the timescale over which inflow occurs, and $A$ is a normalization term. 

The star formation of a galaxy is found to be correlated with its gas content, leading to the so-called star formation law. In this work we simply assume a linear Schmidt law \citep{Schmidt1959},
\begin{equation}
\psi(t)=S\times M_{\rm g}(t),
\end{equation}
where $S$ is the star formation efficiency. Along with the star formation process, massive stars 
die quickly and return metal-enriched gas to the interstellar medium. To model the gas return from the dying stars, it is commonly assumed that stars with masses greater than $1{\rm M_{\odot}}$ die almost instantaneously once formed, and return a constant fraction of their mass to the ISM. Under this assumption, the mass return fraction $R$ for one generation of stars is determined by the initial mass function (IMF) of this stellar population. For a canonical Chabrier IMF \citep{Chabrier2003}, this mass return fraction is $R=0.441$, while for a Salpeter IMF \citep{Salpeter1955} we have $R=0.287$ \citep{Spitoni2017}. It is now generally accepted that the IMF is not universal among galaxies \citep[e.g.][]{Cappellari2012}, but there is still not a well accepted solution for the choice of IMF. In this work we simply adopt $R=0.3$ for all generations of stars without modelling this variation in detail.

For the outflow in galaxies, in most of the previous chemical evolution models \citep[e.g.][]{Arimoto1987}, the outflow strength has been assumed to be proportional to the SFR,
\begin{equation}
\label{eq:outflow}
\dot{M}_{\rm out}(t)=\lambda\psi(t),
\end{equation}
where the parameter $\lambda$ is a dimensionless quantity that characterises the outflow strength relative to the star formation activity, and is often called the `wind parameter'. 
However, \cite{Lian2018mzr} found that a constant value of $\lambda$ cannot explain the mass--metallicity relation of both gas and stars in galaxies, so some variation with time is required.  The physical factor we have not included thus far is that wind has to escape the galaxy, and as its mass builds up its potential well will eventually become deep enough to prevent such escape.  As the simplest possible model of this process, we introduce a variable, $t_{\rm cut}$, to parameterize this time dependence in a straightforward way: in the early Universe, when the look-back time $t>t_{\rm cut}$, the galaxy experiences an outflow characterised by the equation above, while at $t<t_{\rm cut}$ the outflow is suppressed.

With all the previous assumptions, the evolution of the gas mass in the model becomes
\begin{equation}
\label{eq:massevo}
\dot{M}_{\rm g}(t)=
 \left\{\begin{array}{lr}
   A e^{-(t-t_0)/\tau}-S(1-R+\lambda) M_{\rm g}(t) &\text{(for $t>t_{\rm cut}$)}\\
   A e^{-(t-t_0)/\tau}-S(1-R) M_{\rm g}(t) &\text{(for $t<t_{\rm cut}$)}
   \end{array}.
   \right.
\end{equation}

To model the chemical evolution of this gas component, we adopt a simple instantaneous mixing approximation, which assumes that the gas in a galaxy is always well mixed during its evolution. In this approximation, the evolutionary picture is clear: inflow brings in gas, star formation occurs and locks a fraction of metals into stars, some stars die immediately thereafter and eject a metal-enriched gas component, which is mixed instantaneously with the rest of the gas in the galaxy, and the outflow blows out a fraction of this chemically-enriched gas. For each generation of stars, a yield parameter $y_Z$ is conventionally used to characterise the
enrichment of metals, which is defined as the fraction of metal mass generated per stellar mass. The value of $y_Z$ could, in principle, be determined from stellar evolution theory, but uncertainties in mass loss in the late stages of stars' lives mean that it is not currently well constrained, so here we leave it as an adjustable parameter in the model. The equation that characterize the chemical evolution can then be written as
\begin{equation}
\dot{M}_{Z}(t)=Z_{\rm in}\dot{M}_{\rm in}(t)-Z_{\rm g}(t)(1-R)\psi(t)+(1-R)y_Z\psi(t)-Z_{\rm g}\dot{M}_{\rm out}(t),
\end{equation}
where $Z_{\rm g}(t)$ is the gas phase metallicity and $M_{Z}(t)\equiv M_{\rm g}\times Z_{\rm g}$. On the right-hand side of this equation, the first term is the inflow, with $Z_{\rm in}$  the metallicity of the infalling gas. It is conventionally assumed that the infalling gas is pristine ($Z_{\rm in}$=0, \citealt{Tinsley1980}). We therefore adopt this assumption, and the first term is neglected in what follows. The second and third terms show the effects of star formation and metal enrichment due to dying stars, and the last term is the effect of the outflow. Using the equations that characterise the gas infall, star formation and outflow, we obtain as the final equations of the chemical evolution
\begin{equation}
\begin{aligned}
\dot{M}_{Z}(t)= S(1-R)(y_Z-Z_{\rm g}(t)) M_{\rm g}(t)
 -Z_{\rm g}(t)\lambda SM_{\rm g}(t),
\end{aligned}
\end{equation}
 and 
\begin{equation}
\dot{M}_{Z}(t)=S(1-R)(y_Z-Z_{\rm g}(t)) M_{\rm g}(t),
\end{equation}
for $t>t_{\rm cut}$ and  $t<t_{\rm cut}$, respectively.

\begin{figure}
    \centering
    \includegraphics[width=0.4\textwidth]{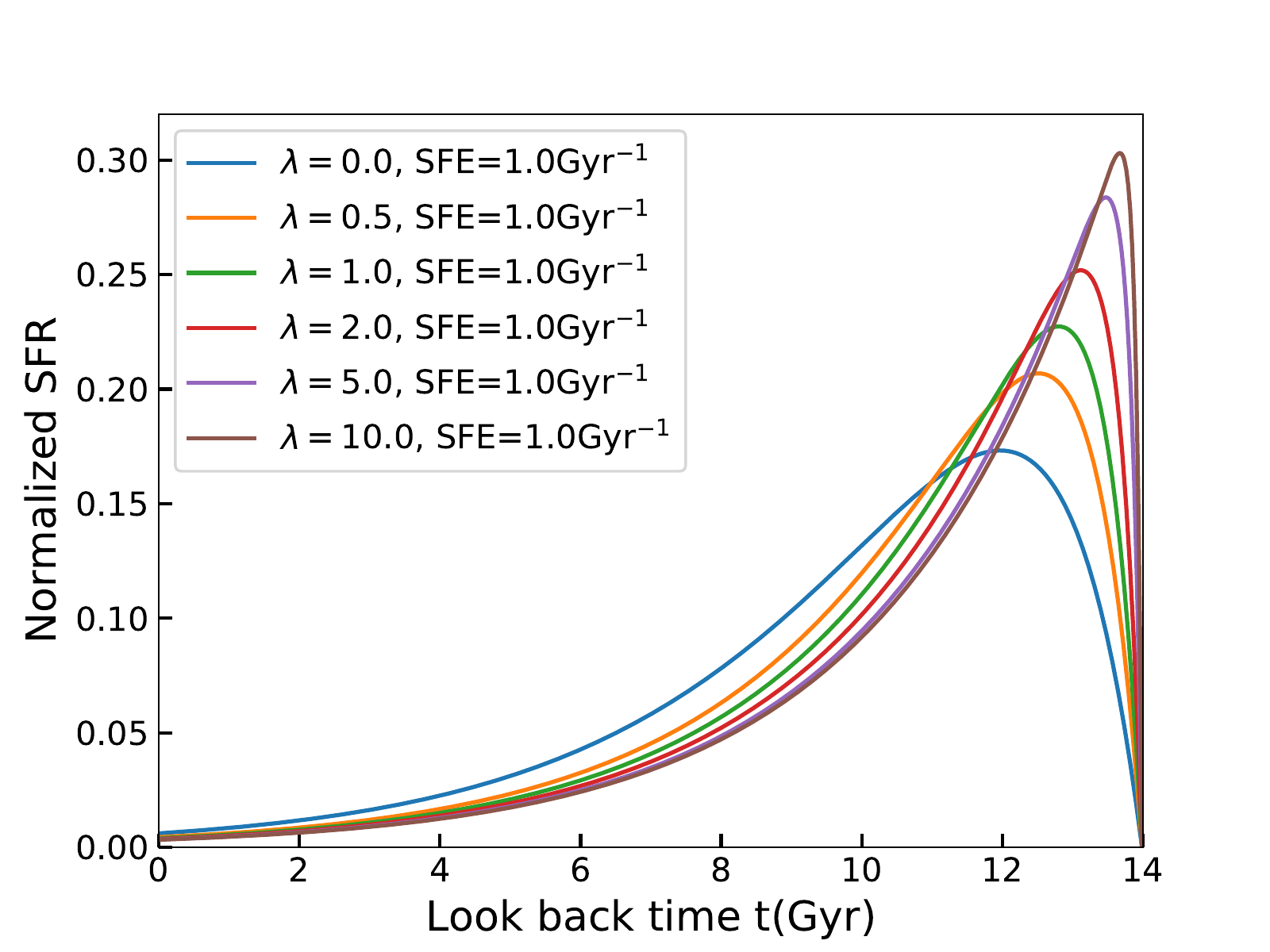}\\
    \includegraphics[width=0.4\textwidth]{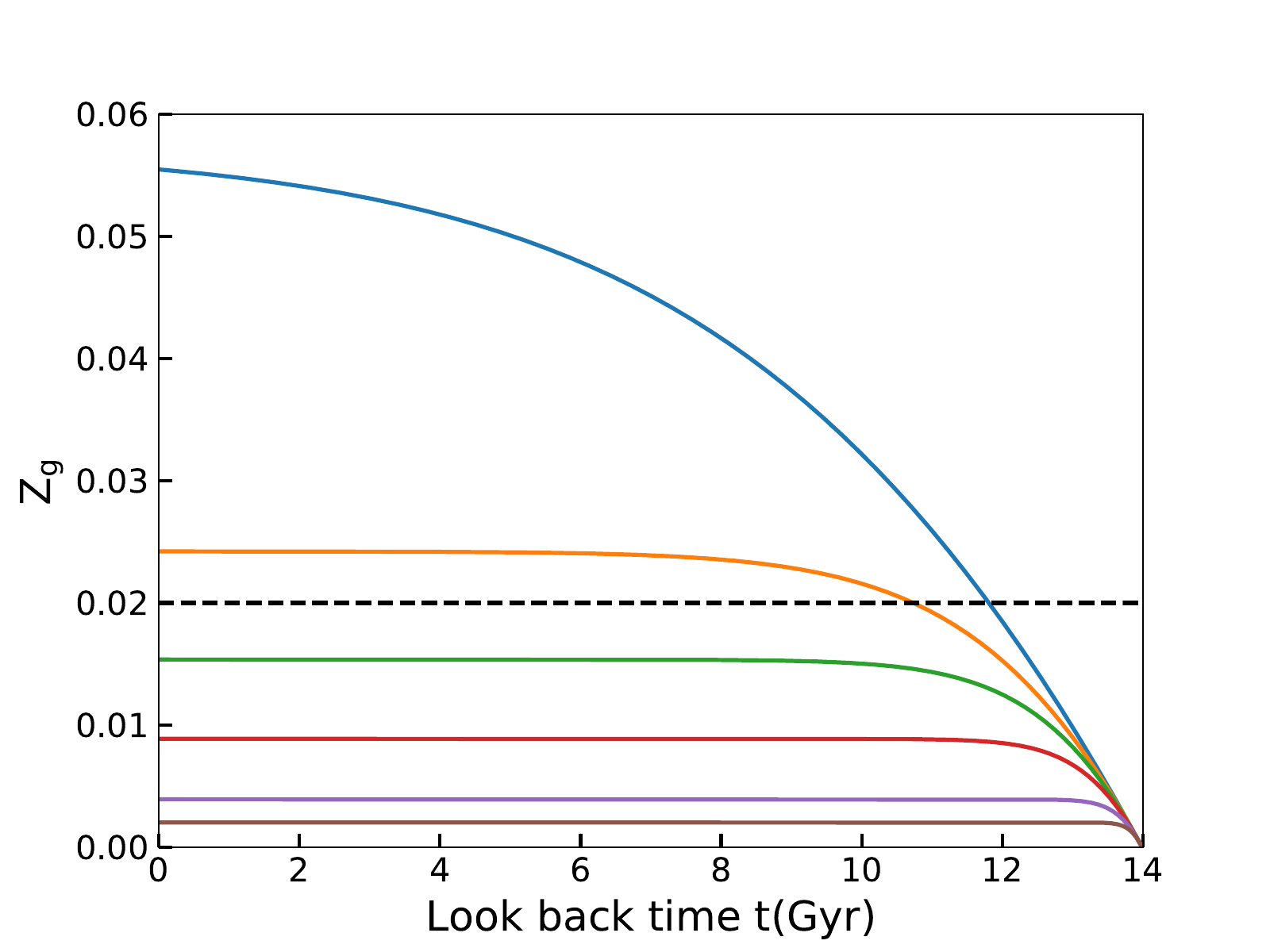}
     \caption{The SFHs (top) and metallicity enrichment histories (bottom) calculated from our chemical evolution model. The model assume an exponentially decaying gas infall starting at the beginning of the universe (assumed to be 14Gyrs ago), with a timescale of $\tau=$3 Gyr. The SFHs are normalized so that the total stellar mass formed is 1M$_{\odot}$. Results with different outflow strengths, characterised by the wind parameter $\lambda$, are shown with different colors as indicated.}
     \label{fig:model_example_outflow}
\end{figure}

\begin{figure}
    \centering
    \includegraphics[width=0.4\textwidth]{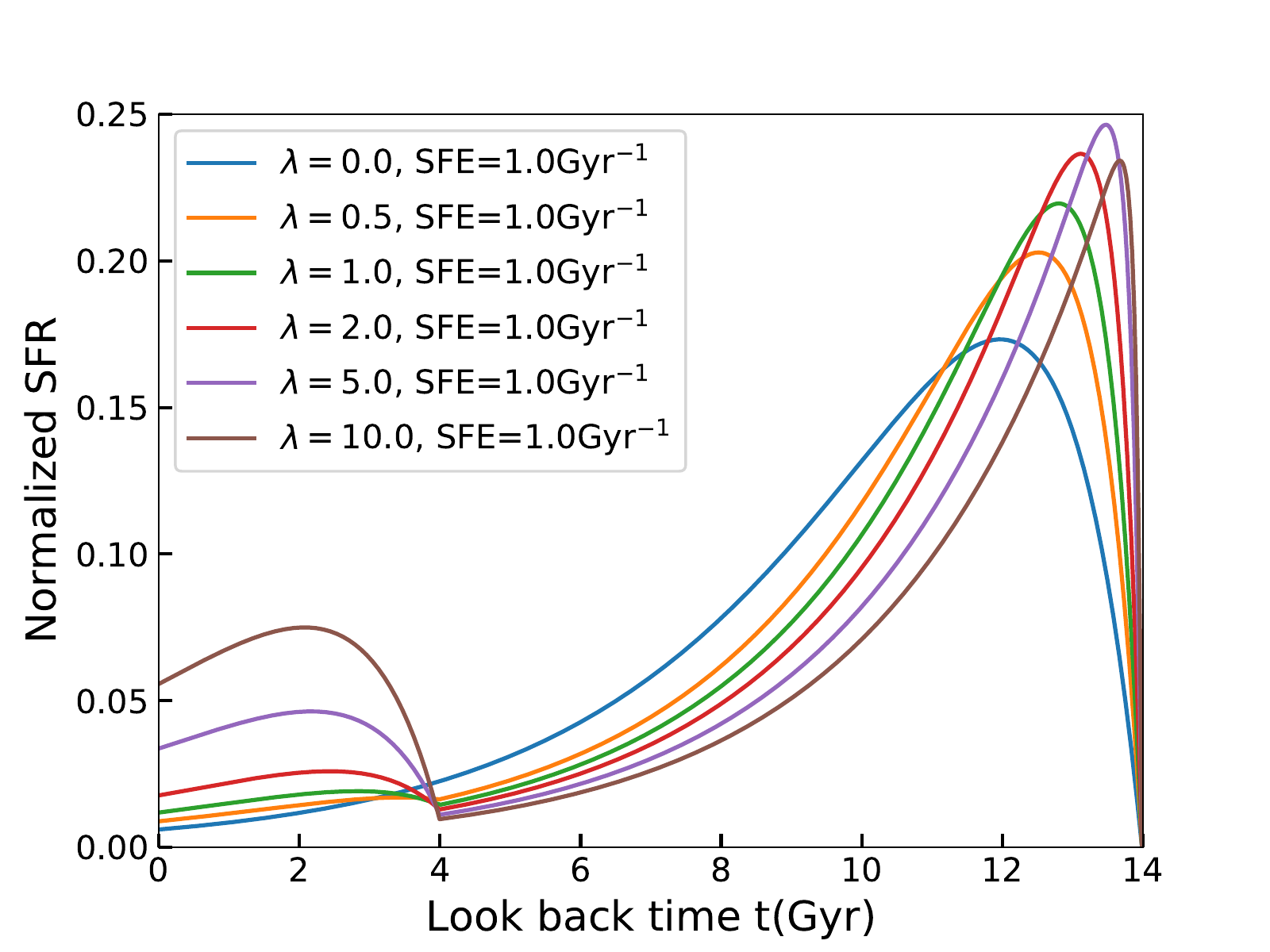}\\
    \includegraphics[width=0.4\textwidth]{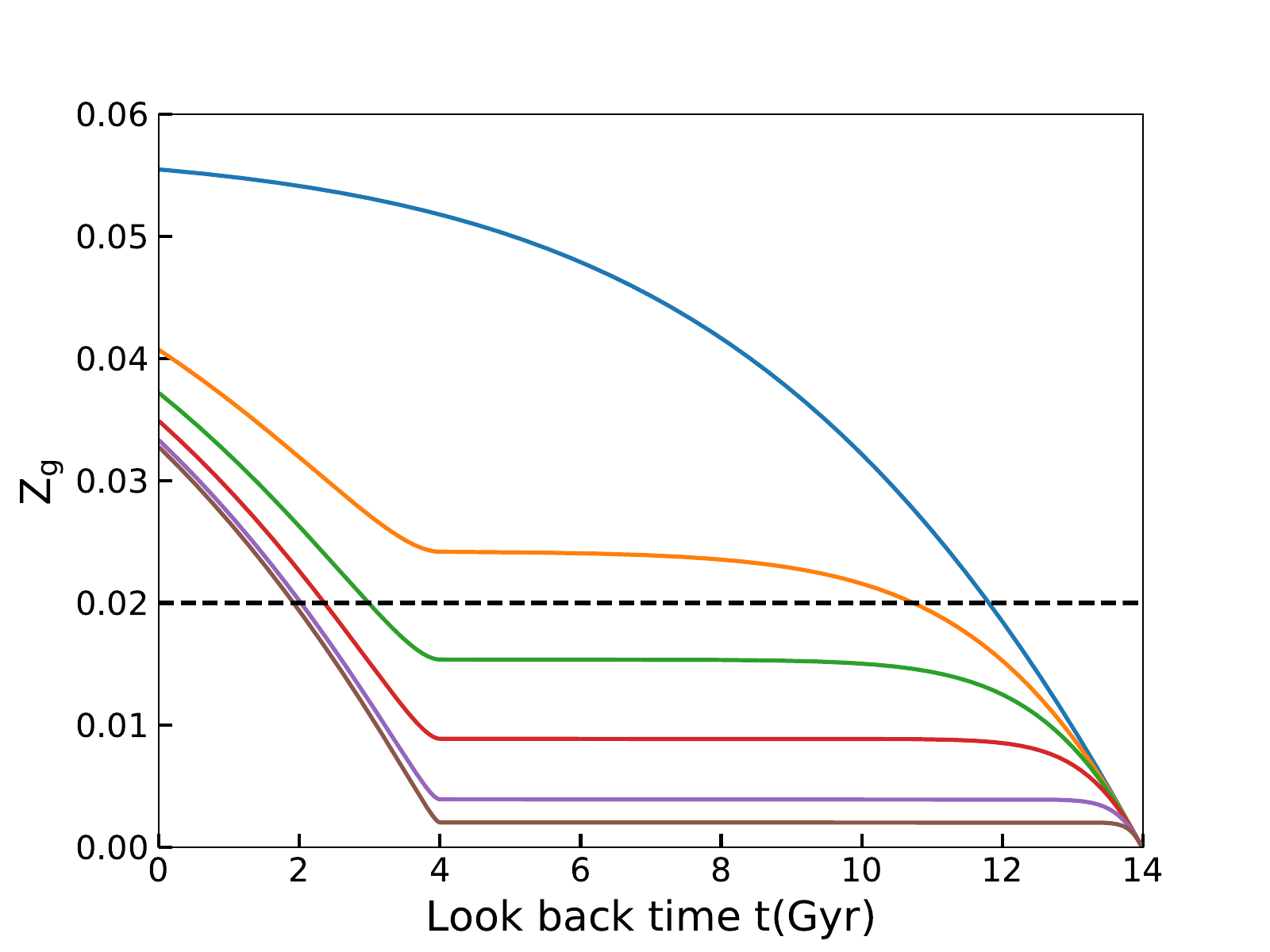}
     \caption{As for \autoref{fig:model_example_outflow}, but with the outflow turned off 4 Gyrs ago.}
     \label{fig:model_example_outflow_turnoff}
\end{figure}

\subsubsection{Exploring the parameter space \& degeneracy analysis}
\autoref{fig:model_example_outflow} shows examples of the ChEHs and SFHs calculated from this chemical evolution model using different model parameters. In the plot we assume an exponentially-decaying gas infall event starting at the beginning of the Universe (assumed to be $14\,{\rm Gyrs}$ ago), with a timescale of $\tau=3\,{\rm Gyr}$. Models were calculated with a constant star formation efficiency $S=1.0$\,Gyr$^{-1}$ and effective yield $y_Z=0.03$.These values characterise the typical evolution of a Milky-Way-like galaxy, and can be readily compared with models presented in \cite{Spitoni2017}. Under such assumptions, we naturally obtain a decaying SFH in the absence of outflows, with the metallicity of the galaxy steadily increasing with time (blue line). The inclusion of the outflow (other coloured lines) blows away material, leading to shorter star-formation timescales. In addition, the outflow  blows away metals, which suppresses the chemical enrichment process in the galaxy. For galaxies with strong outflows, the gas metallicity would increase in the beginning and then stay constant after a balance is reached, with the final metallicity strongly correlated with the outflow strength. In the models presented in  \autoref{fig:model_example_outflow_turnoff}, the outflows stop at $4\,{\rm Gyrs}$ ago, which triggers secondary star formation events seen in their SFHs (top panel). In addition, this secondary star-formation induces a relatively rapid metallicity enrichment process (bottom panel).

\begin{figure}
    \centering
    \includegraphics[width=0.4\textwidth]{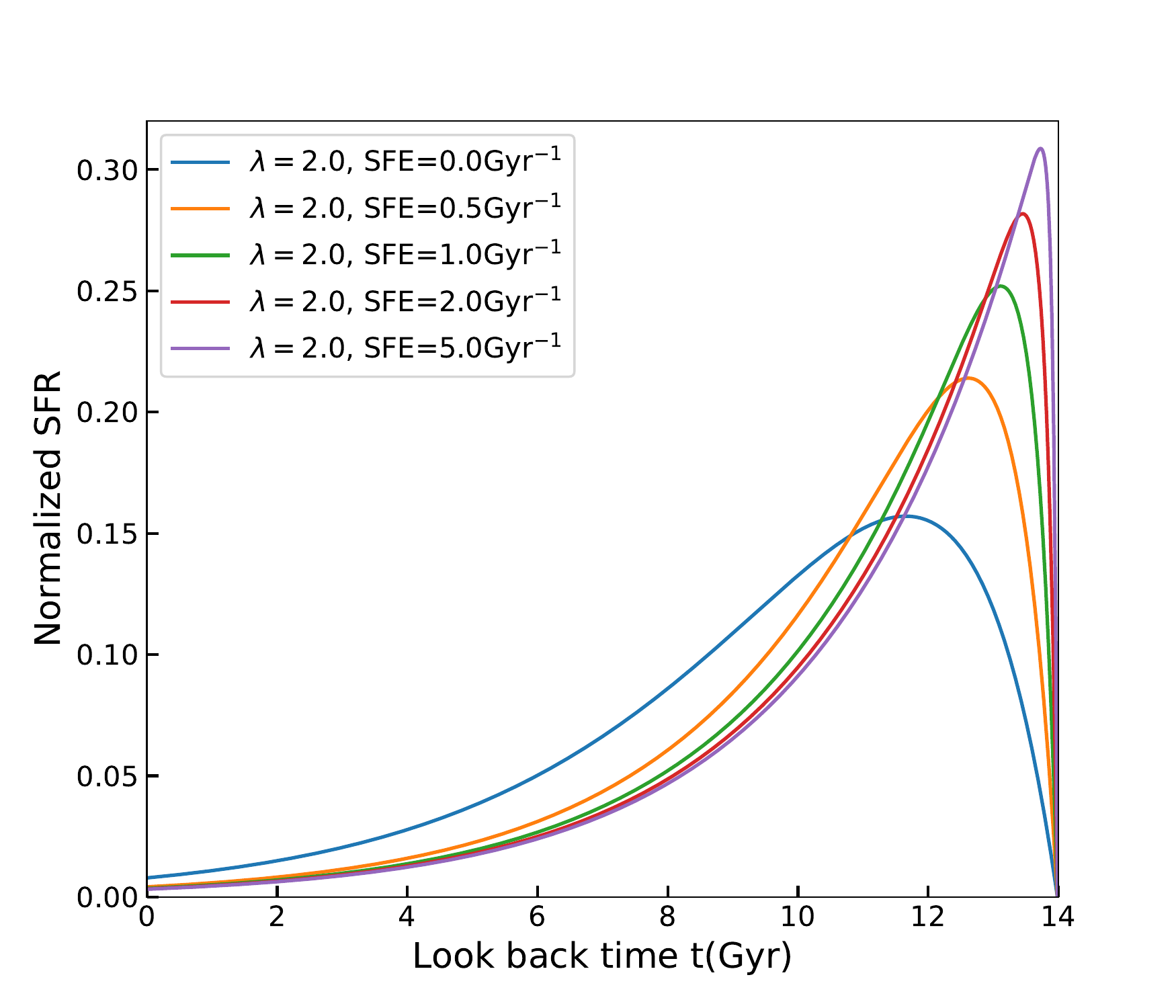}\\
    \includegraphics[width=0.4\textwidth]{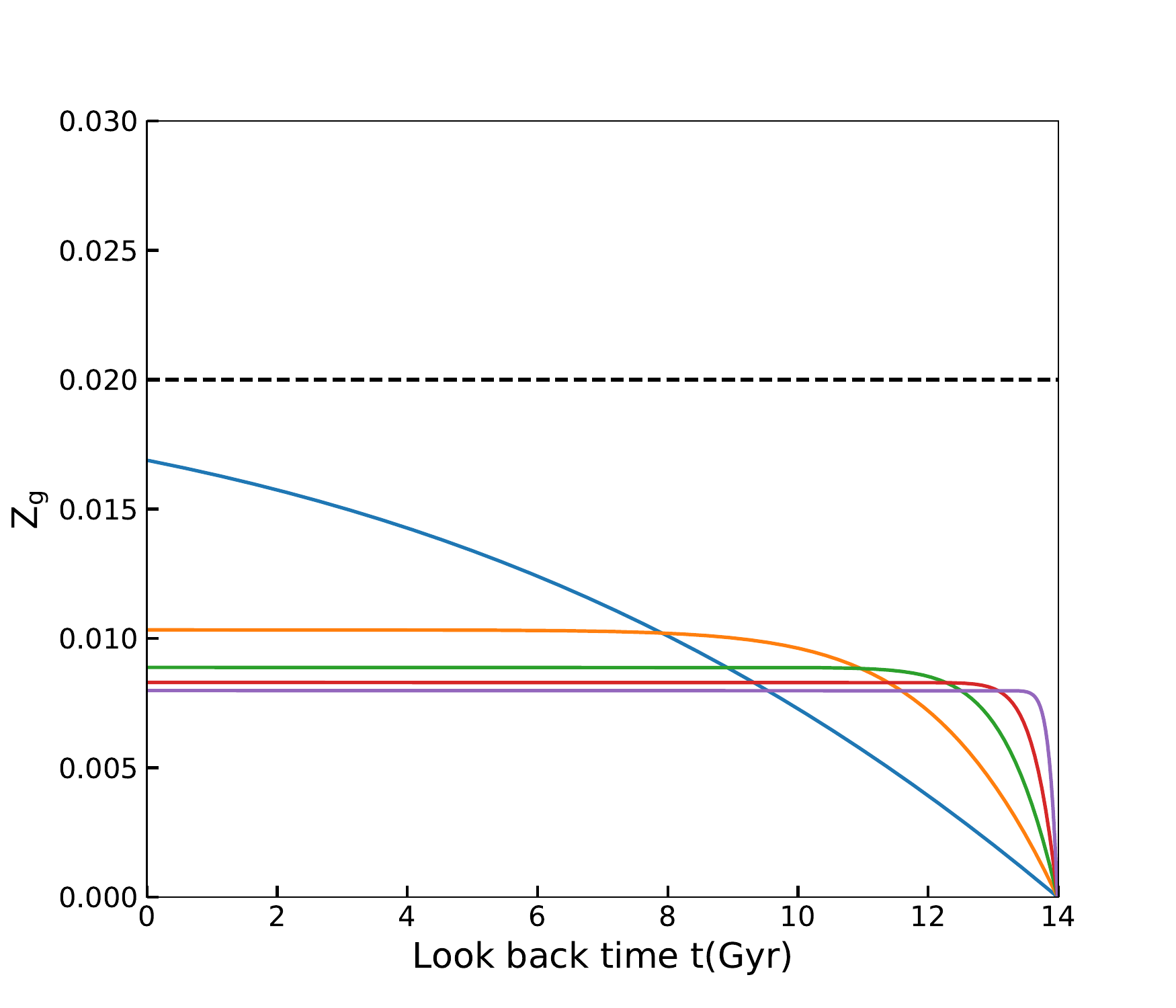}
     \caption{Similar to \autoref{fig:model_example_outflow}, but with fixed wind parameter $\lambda$ and varying SFE, as indicated.}
     \label{fig:model_example_eff}
\end{figure}

\begin{figure}
    \centering
    \includegraphics[width=0.4\textwidth]{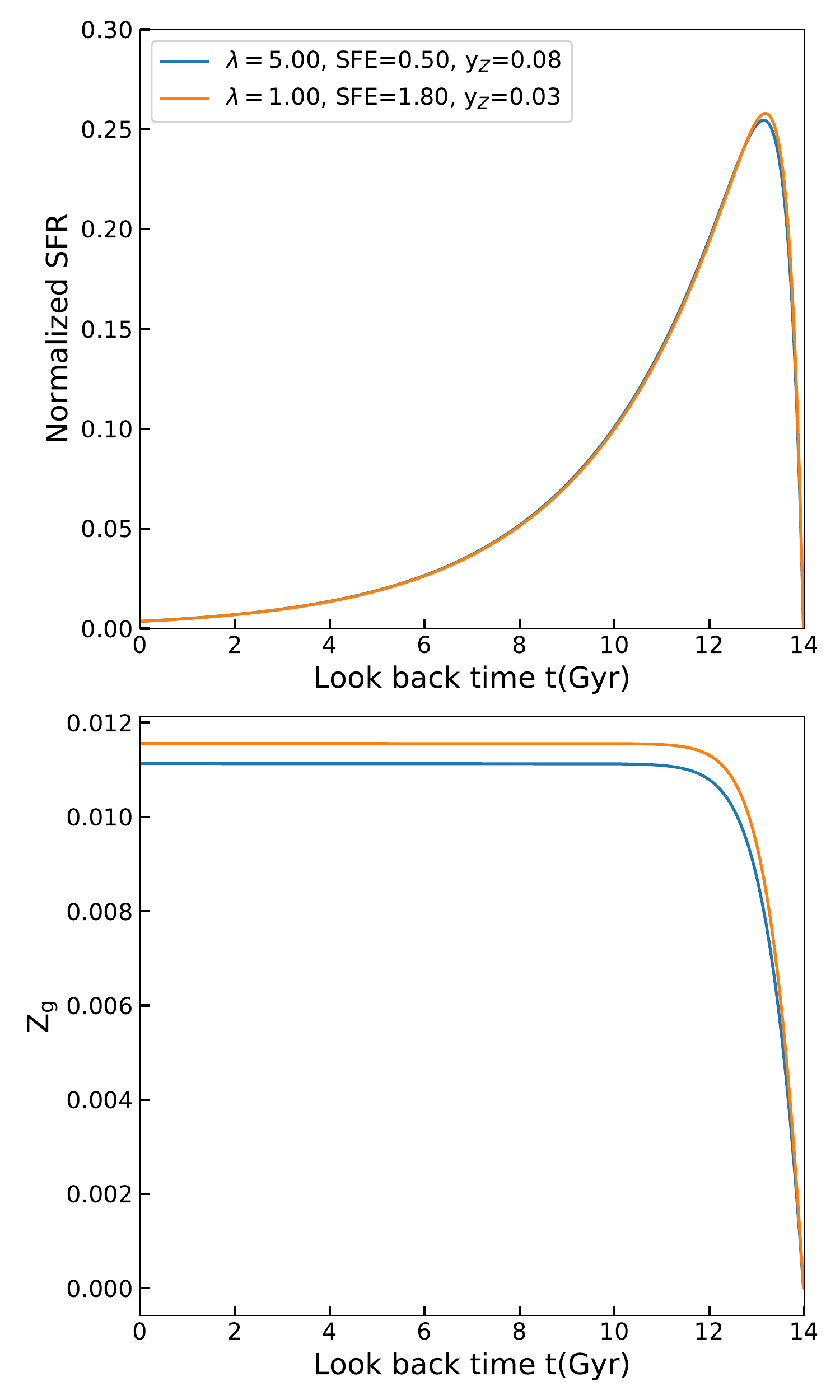}\\
     \caption{An example showing the degeneracy of model parameters. Two different sets of model parameters predict almost identical SFH (top) and ChEH (bottom).}
     \label{fig:model_example_degen}
\end{figure}

For comparison, \autoref{fig:model_example_eff} shows examples similar to those presented in \autoref{fig:model_example_outflow}, but with a constant wind parameter of $\lambda=2.0$ and varying SFE. It can be clearly seen from the top panel that the changing of SFE and wind parameter have vary similar effects on the SFHs predicted by the model. Galaxies with high SFEs would have short timescales in their star formation, which cannot be distinguished from models with strong outflows. This degeneracy can be straightforwardly understood from \autoref{eq:massevo}, in which S and $\lambda$ work in similar ways. Differences are expected to be seen in their metallicity evolution (bottom panel): as galaxies with high SFEs but weak outflows do not lose their metals during their evolution, those galaxy would accumulate their metals faster and would have higher final metallicities when the final equilibrium is reached. However, as the fast enrichment process happened at the star burst phase in the early Universe, the timescale of the metallicities enrichment process is not likely to be well constrained using the available spectral data.
Moreover, by increasing the effective yield, galaxies with strong outflows can also reach a high final metallicity, and the models can then become totally degenerate. We show such an example in \autoref{fig:model_example_degen}, which shows that the two different sets of model parameters predict almost identical SFH (top panel) and ChEH (bottom panel), which we cannot hope to distinguish in the fitting process.

This analysis indicates that in real fitting we cannot leave all three parameters -- the SFE, wind parameter and effective yield -- to be free.
Instead, some kind of auxiliary information that can help us to pin down one of them is crucial in correctly constraining the mass and metallicity evolution in galaxies. Among these parameters, the outflow property varies with galactic scale physics, such as the depth of the potential well and AGN activity, which cannot be directly inferred from available information. The yield in a galaxy correlates with its IMF, which has been found to vary between galaxies in a manner that is still poorly understood. In addition, the effective yield can be affected by the uncertainties in the returned mass fraction of stars, and even uncertainties in the instantaneous mixing approximation, and thus may deviate from any known theoretical values. In contrast, however, many studies have shown that star formation efficiency is largely dictated by measurable local properties 
\citep{Leroy2008,Shi2011}. For example, \cite{Leroy2008} measure the SFE in 23 nearby galaxies and found that the SFE of molecular gas is almost constant, and that the ratio between H{\sc i} and H{\sc ii} is tightly correlated with the local stellar mass surface density. Inspired by such a correlation, \cite{Shi2011} proposed the extended Schmidt law, in which the SFE depends explicitly on the stellar mass surface density ($\Sigma_{*}$) via 
\begin{equation}
\label{eq:sfe}
    S (yr^{-1})=10^{-10.28\pm0.08} \left( \frac{\Sigma_*}{M_{\odot} pc^{-2}} \right).
\end{equation}
Given that this quantity is relatively tightly constrained by local measurable astrophysics, we choose to lift the degeneracy by fixing the SFE using this empirical calibration, and leave the outflow strength and effective yield as the remaining free parameters. We use \autoref{eq:sfe} to estimate the SFE of the galaxy, obtaining an approximate average value for the stellar mass surface density from the values of mass and effective radius in the NASA-Sloan Atlas $\Sigma_*=2\times M_*/(\pi R_e^2)$. In principle, we should allow for the fact that this characteristic surface density will evolve with time as both $M_*$ and $R_{\rm e}$ of a galaxy increase as it grows. However, by using the mass and size evolution estimated from MaNGA spiral galaxies \citep{Peterken2020}, we find that the average stellar mass surface density within 1$R_{\rm e}$ does not change dramatically in the last $10\,{\rm Gyr}$, leading to less than a $10\%$ variation in the predicted SFE. This time evolution has a negligible effect on the derived SFH and ChEH, and we thus neglect it in the modelling process.

In summary, this model with an exponentially decaying gas infall and constant level of outflow leads to a steadily decaying SFH, with metallicities quickly increasing in the beginning and remaining constant after the equilibrium state is reached. If the outflow turns off at some point, the galaxy experiences a secondary star formation event associated with a more rapid metallicity enrichment process. The three key parameters,  the SFE, wind parameter and effective yield, are strongly degenerate with each other and cannot all be left as free parameters. Fortunately, however, we have good physical motivation that allows us to fix one of these parameters, the SFE, using independent auxiliary data, allowing the two remaining parameters to be derived.

\subsection{Spectral Analysis approaches}
We constrain the chemical evolution of galaxies through fitting their optical spectra, gas phase metallicities, and current star formation rates provided by MaNGA. The fitting is achieved using an updated version of {\tt BIGS}, Bayesian Inference of Galaxy Spectra, which is a Python spectral fitting code developed in our previous work \citep{Zhou2019}. In this section we describe the fitting process in some detail.

{\tt BIGS} utilizes the {\tt MULTINEST} sampler \citep{Feroz2009,Feroz2013} and its \textsc{Python} interface \citep{Buchner2014} to sample the posterior distributions. To begin with, a set of parameters are generated using an appropriate prior distribution (see \autoref{tab:paras}). We use the chemical evolution model above to predict the SFH and ChEH from those parameters. The present-day star formation rate and gas phase metallicity $Z_{\rm g}$ are obtained from the SFH and ChEH, respectively. To model the spectra, we adopt the canonical stellar population synthesis (SPS) method. In the SPS approach, using knowledge of stellar evolution, stars in the galaxies are divided into different groups of specific ages and metallicities, which are often referred to as single stellar populations (SSPs). The spectra of these populations are calculated with stellar spectra templates, and are often provided as SSP models  \citep[e.g.][]{BC03,Maraston2005,Vazdekis2010,Maraston2020}. After convolving with the SFH and ChEH given above, composite stellar populations (CSPs) can be calculated, which can then be compared with observations. In this work we use the high-resolution \citet{BC03} models, which are constructed with the STELIB empirical stellar templates \citep{Borgne2003}. These high spectral resolution SSPs cover the wavelength range from 3200{\AA} to 9500{\AA} with a FWHM resolution of 3\AA. We use the model calculated with the ‘Padova1994’ stellar evolution model \citep{Bertelli1994}, which covers metallicities from $Z =0.0001$ to $Z = 0.05$, and ages from $0.0001\,{\rm Gyr}$ to $20\,{\rm Gyr}$ (but only those with an age of less than $14\,{\rm Gyr}$ are included in this analysis). In this work we analyze spectra observed in the MaNGA survey (see \autoref{sec:data}). Since the wavelength range of the original MaNGA data is $3600$ -- $10300$\AA, all at low redshifts, these templates are well matched to the rest-frame wavelength range of the observations. This model is thus well-suited to fitting MaNGA spectra. The same models have also been widely used in spectral analysis works over the past 20 years, which means that this choice is appropriate comparing our results with many canonical works. To account for dust attenuation effects, a simple screen dust model characterized by the Calzetti attenuation curve \citep{Calzetti2000} is added to the final spectral fit. To compensate for the resolution differences between the model spectra and MaNGA data, due to both kinematic and instrumental broadening effects, we fit the observed spectra with pPXF using the BC03 templates to derive an effective velocity dispersion. The model CSP are convolved with this effective velocity dispersion to match their resolution similar to the MaNGA data. Finally, the model spectra and model gas phase metallicity 
are compared with the observed spectrum to determine how well it fits.

As a measure of goodness of fit, we use a $\chi^2$-like likelihood function,
\begin{equation}
\label{likelyhood}
\ln {L(\theta)}\propto-\sum_{i}^N\frac{\left(f_{\theta,i}-f_{\rm D,i}\right)^2}{2f_{\rm err,i}^2}-
\frac{(Z_{\rm g,\theta}-Z_{\rm g,D})^2}{2\sigma_{Z}^2
}-\frac{(\psi_{0,\theta}-\psi_{\rm 0,D})^2}{2\sigma_{\psi}^2},\,
\end{equation}
where $f_{\theta, i}$ is the flux at the $i$-th wavelength point 
as predicted for our model specified by the parameter set $\theta$, $f_{\rm D, i}$ is the 
flux at the same wavelength in the stacked spectrum, and $f_{\rm err,i}$ is the corresponding error spectrum, $N$ is 
total number of wavelength points. $Z_{\rm g,\theta}$ and $Z_{\rm g,D}$ are the current gas phase metallicities of the galaxy from our model predictions and MaNGA observations respectively, with $\sigma_{Z}$ being the uncertainty of the gas phase metallicity estimates. Similarity, $\psi_{0,\theta}$ and $\psi_{0, \rm D}$ are the current SFR of the galaxy from our model predictions and MaNGA emission-line observations respectively, and $\sigma_{\psi}$ characterise the uncertainty of the SFR data.

Determining appropriate values for $\sigma_{Z}$ and $\sigma_{\psi}$ is not entirely straightforward.
Compared with traditional fitting that only include spectra such as \cite{Zhou2019}, the likelihood function of \autoref{likelyhood} combines the constraints from stellar continua and absorption lines with those from emission lines. As there is no prior knowledge about which part is more important in constraining the evolution of the galaxy, one would expect to have an additional normalization parameter to adjust the relative importance of the different parts in constraining the fit. However, adding normalization parameters to the last two terms in \autoref{likelyhood} is mathematically indistinguishable from changing the values of $\sigma_{Z}$ and $\sigma_{\psi}$. In this work, we do not try to add any additional normalization, but just use empirical uncertainties. The typical intrinsic variation for gas metallicity from the $O3N2$ index is found to be around 0.14dex \citep{PP04}. Compared with the small  statistical uncertainty ($\lesssim 2\%$), the uncertainty in the gas phase is dominated by systematic variations and cannot be easily estimated. We therefore simply use $\sigma_{Z}=0.1Z_{g,D}$, which 
is a compromise between the constraints from stellar and gas components. For 
the SFR, as the dust attenuation effect that can affect the measurements of the H${\alpha}$ flux have been corrected, the uncertainties mainly come from the conversions from H${\alpha}$ luminosity to SFR. The factor used in \autoref{eq:sfr} is directly calculated from stellar population syntheses models, which would be expected to have systematic uncertainties similar to the model spectra. To model this uncertainty in detail goes beyond the scope of this work. As the uncertainties of SPS models are also not explicitly included in the first term of \autoref{likelyhood}, we do not take into account this uncertainty in SFR. We thus only assign the typical uncertainty in the measurement of  H${\alpha}$ flux ($2\%$; see \autoref{sec:data_re}) to the uncertainty of SFR  so that $\sigma_{\psi}=0.02\psi_{0,D}$. We note that the appropriate treatment of these uncertainties is not uniquely defined. In this proof-of-concept work we seek to use as much information as possible to derive reliable results.  Through extensive testing, we have found that the above choices make good use of the information from the gas phase metallicity and the SFR, while not significantly altering the fitting quality of the optical spectrum. In fact, given the large number of wavelength points for the absorption-line spectra, the major constraints to the model come from the spectra themselves, so that most of the fitting results are not greatly affected by the emission-line constraints. The inclusion of current gas phase metallicity helps to determine the current metal content of galaxies that do not have much recent star formation and thus cannot be properly constrained by the absorption-line spectra alone. Removing the gas phase metallicity constraints is found to lead to unphysically extreme values of metallicity in such galaxies. The current SFR constraint, by contrast, has most effect in galaxies that have experienced strong recent star formation, which cannot be easily characterised by the continuous model used in this work, so are leveraged by this constraint. As noted, these choices are not unique, but represent a calibrated compromise that makes physical sense for this data set; other calibrations may be appropriate for different data sets, but then some care needs to be taken when comparing results.

Once the likelihood is calculated, the MULTINEST sampler updates the posterior probability accordingly and generates a new parameter set, until a convergence criterion is reached. After convergence, {\tt BIGS} outputs the posterior distributions of the model parameters and the Bayesian evidence, which will be used in the subsequent analysis. In selecting the prior range for the parameters, theoretical and commonly-used values are considered. For the yield parameter, a canonical Chabrier IMF has $y_Z$=0.0631, while a Salpeter IMF gives $y_Z$= 0.0301 \cite{Spitoni2017}. The prior ranges are set to cover these theoretical values and extend reasonably to account for possible outliers. The prior ranges of inflow starting time and outflow turn-off time are set to be consistent with the age of Universe (assumed to be 14Gyrs for simplicity), with time scales of the inflow ranging from an instantaneous burst (0.0Gyr) to reasonably flat (14.0Gyr). For the  wind parameters, we use the values from \cite{Spitoni2017}, while the dust attenuation values are designed to cover the ranges from stellar colour excesses for MaNGA galaxies provided in \cite{Li2021}. We list all the model parameters and their prior assumed values in \autoref{tab:paras}.

\begin{table}
	\centering
	\caption{Priors of model parameters used to fit galaxy spectra}
	\label{tab:paras}
	\begin{tabular}{lccr}
		\hline
		Parameter & Description & Prior range\\
		\hline
		$y_Z$ & Effective yield & $[0.0, 0.08]$\\
		$\tau$ & Gas infall timescale & $[0.0, 14.0]$Gyr\\
		$t_{0}$ & Start time of gas infall & $[0.0, 14.0]$Gyr\\
		$\lambda$ & The wind parameter & $[0.0, 10.0]$\\
		$t_{\rm cut}$ & The time that outflow turns off & $[0.0, 14.0]$Gyr\\
		$E(B-V)$&  Dust attenuation parameter & $[0.0, 0.5]$\\
		\hline
	\end{tabular}
\end{table}

\section{Tests}
\label{sec:test}

\begin{figure*}
    \centering
    \includegraphics[width=1.0\textwidth]{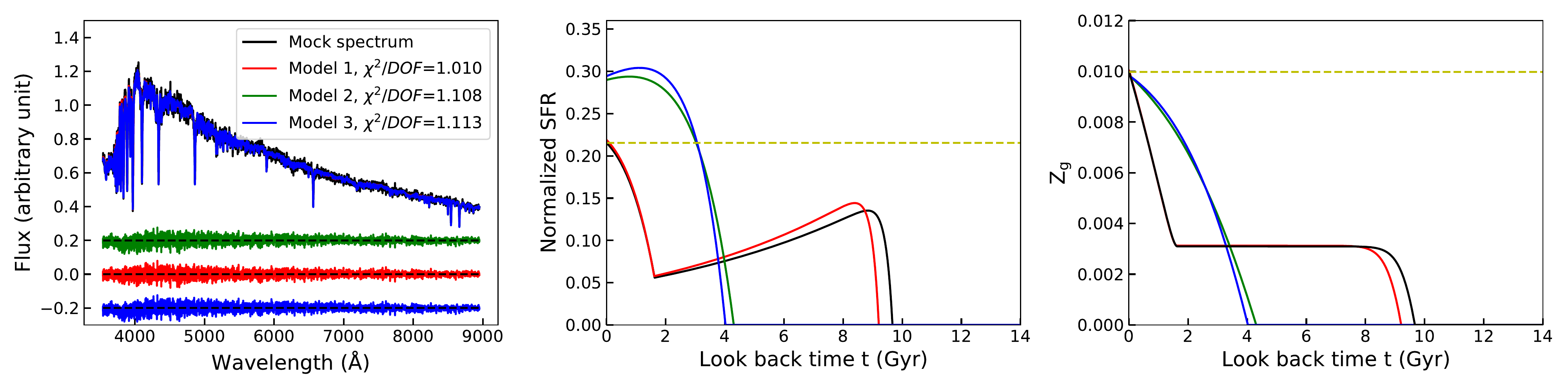}\\
     \caption{An example of fitting the mock spectrum with three models. 
     The black line in the left panel shows the mock spectrum, which is calculated through our chemical evolution model with a set of randomly generated parameters. The Corresponding SFH and ChEH from those parameters are shown as black lines in the middle and right panels respectively. The yellow horizontal dash lines in the middle and right panels mark the current SFR and gas phase metallicity used in constraining the model. Best-fit results from the three models (see \autoref{sec:test}) are shown with three different colors, with values of $\chi^2$ per degree of freedom (DOF) showing at top-right of the left panel to quantify the residuals.
     }
     \label{fig:example_mock}
\end{figure*}

\begin{figure*}
    \centering
    \includegraphics[width=1.0\textwidth]{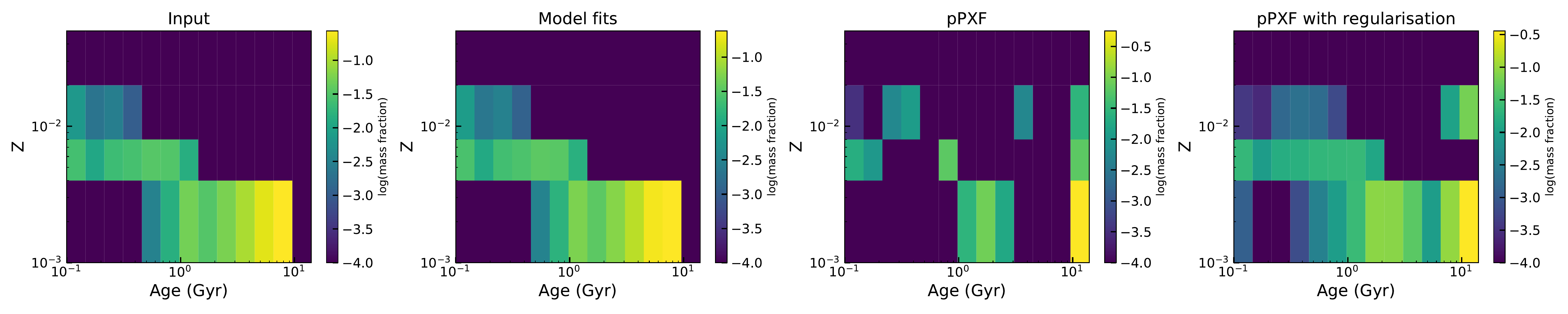}\\
     \caption{Comparison between our model fits with {\tt pPXF}. Plots show the mass fractions of SSPs in the metallicity-age space.From left to right, 
     the first panel plots the input model that is used to calculate the mock spectrum, corresponding to black lines in \autoref{fig:example_mock}; best-fit results recovered using model 1 (red lines in \autoref{fig:example_mock}) are shown in the second panel; the third panel comes from {\tt pPXF} fitting of the mock spectrum without regularization, while the last panel shows the results from regularized pPXF fits.
     }
     \label{fig:example_mock_ppxf}
\end{figure*}

\begin{figure*}
    \centering
    \includegraphics[width=1.0\textwidth]{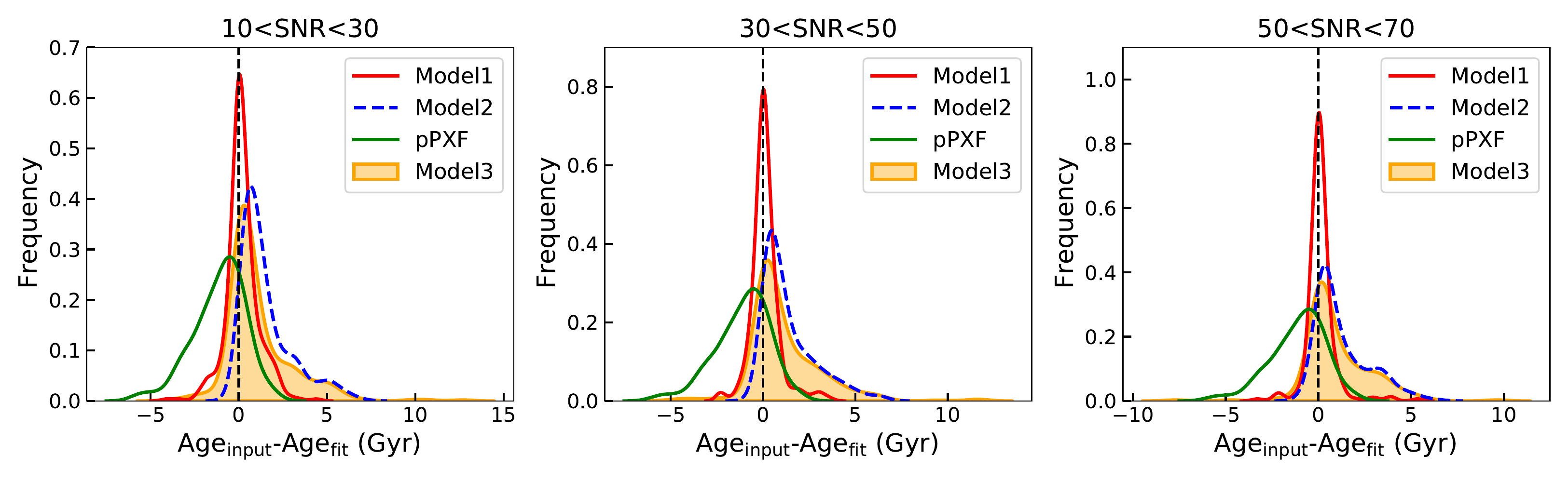}\\
    \includegraphics[width=1.0\textwidth]{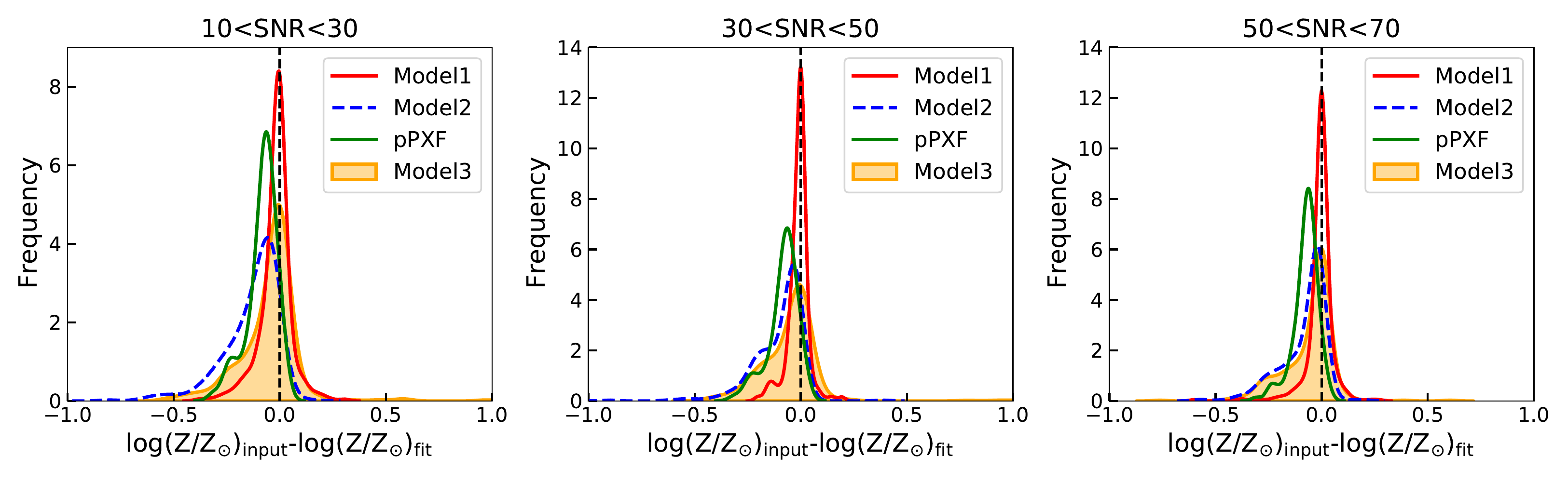}\\
     \caption{The distribution of the difference in the mean age(top) and stellar metallicity(bottom) between the best-fit model and input values. The three columns are results from fitting mock spectra with different levels of SNR, as indicated. Results obtained from the three models and results from pPXF with regularisation are shown with different colors as indicated. Probability distributions are each normalized to 1.}
     \label{fig:mock_stats}
\end{figure*}

We have undertaken a wide range of tests to assess the robustness of this new method, how well the parameters are likely to be constrained by real data, and what biases the results might contain. To this end, as well as varying the amount of freedom in the fitting process, we have tested mock data generated both from our assumed model and from an independent more random model. In this section, we describe the results of these tests.

We first examine whether observed data can realistically provide enough information to constrain a model of this complexity. Accordingly, we generate a simulated data sample using the following procedure. To begin with, a set of model parameters are generated following the prior distribution listed in \autoref{tab:paras}. Those parameters are used to calculate the mock SFH and ChEH using the chemical evolution model. As the currently available SSP templates only cover stellar metallicities up to 0.05, we exclude parameter sets that will predict a current gas phase metallicity larger than 0.05. The valid parameters are sent to {\tt BIGS} to generate a simulated spectrum using the BC03 SSP templates. Gaussian noise is added to the spectrum so that the final SNR falls between 10 and 70. By repeating this process, we generate a mock sample of 1000 spectra, which are then fitted with the procedure described in the last section. To assess whether the model is over-constrained, we carry out this fitting process assuming three forms of chemical evolution model: model 1 is the full input model, which includes time-dependent inflow and outflow processes and will be used throughout this work;  model 2 only considers the inflow process, i.e. the wind parameter $\lambda$ is set to 0 throughout the fitting; model 3 takes the outflow into account, but the outflow strength has no time-dependence, i.e. $t_{\rm cut}=0$ throughout the fitting.

\autoref{fig:example_mock} shows an example of such fitting. From the best-fit spectra shown in the left panel, it can be seen that all the three models do a reasonable job of fitting the observed spectrum. However, the residuals of models 2 and 3 are higher (as shown by the $\chi^2$ per degree of freedom indicated in the plot), so, with spectra at sufficient levels of signal-to-noise, one can distinguish between them. While model 1 recovers well both the SFH and ChEH (red lines in each plot), we see biases in the reproduced SFH and ChEH for model 2 and model 3 in the middle and right panels. The introduction of a time-dependent outflow process makes the model flexible enough to describe two major star formation and chemical enrichment episodes, which cannot be properly characterised by simpler models such as model 2 and model 3. This flexibility is crucial in fitting galaxies that have complex formation histories, making full use of high SNR spectra \citep{Zhou2019}. 

As a further test, we compare our model fits to this mock spectrum with results obtained via a more conventional technique, using {\tt pPXF}. To do so, we fit the mock spectrum in \autoref{fig:example_mock} with {\tt pPXF} and derive the mass fractions of SSPs from the {\tt pPXF} fits. We undertaken two kinds of pPXF fits: one fits the mock spectra without any regularization and the other includes regularization as suggested in \cite{Cappellari2017}. To perform the regularized fit, we first normalized all the SSP templates so that each template has a median of one, and do the same for the mock galaxy spectrum. We then use a regularization strength `regul=100' in the pPXF fitting \citep{Cappellari2017}. As we did not add any effect of dust attenuation and the SSP libraries used in generating and fitting the mock spectra are exactly the same, we
do not expect any continuum differences to require correcting, and thus no additional polynomials are used during the fitting process. From left to right, the mass fractions of SSPs from the SFH and ChEH of the mock inputs, model 1 fitting results and {\tt pPXF} results without and with regularization are shown in \autoref{fig:example_mock_ppxf}. While in general the {\tt pPXF} fit matches the input data, it is clear that its inability to impose physical smoothness on the two-dimensional fit leads to significant biases and noise. 
The introduction of a regularization term helps to reach a balance between the fitting quality and the smoothness of the derived physics. But as the smoothness has no physics behind it, some unexpected populations such as the relatively metal-rich old population ($\sim$10Gyr) are seen even in the regularised {\tt pPXF} results. In addition, as the gas phase metallicities cannot be included in the fitting processes of {\tt pPXF}, the youngest stellar populations are not well constrained. In this case, for example, we see a weak but very metal-poor young population ($\sim0.1\,{\rm Gyr}$) in the {\tt pPXF} results, which is unphysical given the current gas phase metallicity in this particular simulated galaxy.

To statistically compare the true values to the recovered physical properties, we calculate the mass-averaged age and metallicity from the SFH and ChEH. The differences of ages and metallicities between the input model and best-fit results of the three models are shown in \autoref{fig:mock_stats}. It can be seen that when the SNR is low (<30), all three models get average ages and metallicities with large scatter when compared to the input values. Model 1 (red) reproduces the input values slightly better than model 2 (blue) and model 3 (orange) in the sense that the results are less biased. This is not unexpected as the data are simply not of high enough quality to constrain the model in detail. When the SNR increases, the best-fit results from model 1 converge to the input model, while results from model 2 and 3 are still significantly biased. These results indicate statistically that the adopted method is able to recover the input model at SNR>30, and shows clear advantages compared to simpler models.  As comparison, we also show in \autoref{fig:mock_stats} the results derived using {\tt pPXF} with regularization (green line). It is seen that, since it is exactly the input model, Model 1 always behaves better than {\tt pPXF}. With models 2 and 3, pPXF does a comparable job in reproducing the average ages, while doing better in recovering metallicities. This is expected as {\tt pPXF} has more flexibility to fit any combinations of SFH and ChEH so that the inputs can be reasonably reproduced. However, the lack of any physical link between the SFH and ChEH also makes it possible for {\tt pPXF} to derive a solution different to the input model, especially in the low signal-to-noise ratio cases.

\begin{figure*}
    \centering
    \includegraphics[width=1.0\textwidth]{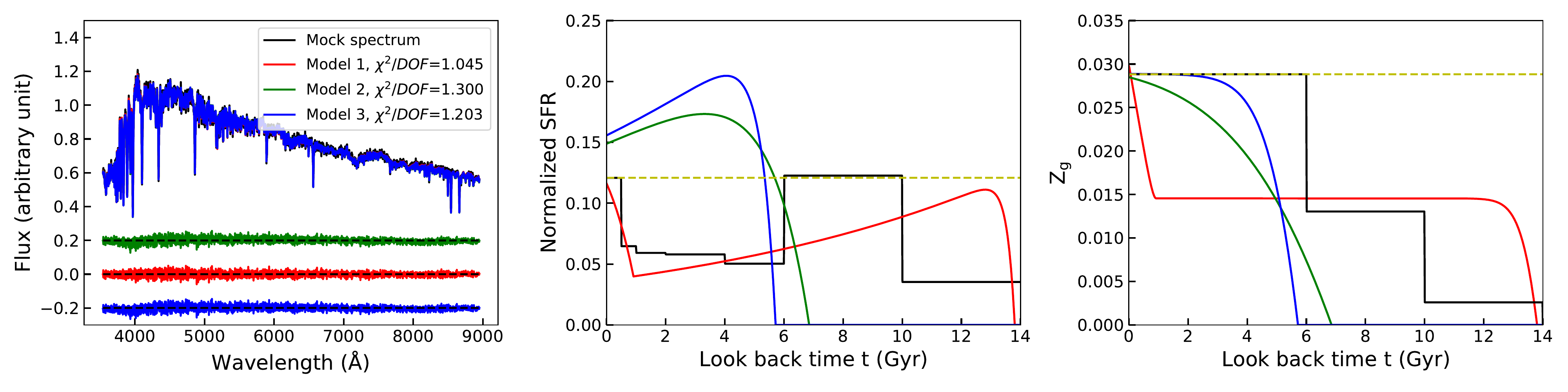}\\
     \caption{An example of fitting a mock spectrum with three models. 
     The black line in the left panel shows the mock spectrum, which is calculated through assuming a stepwise SFH and ChEH shown as black lines in the middle and right panels respectively. The yellow horizontal dash lines in the middle and right panels mark the current SFR and  gas phase metallicity used in constraining the model, respectively. Best-fit results from the three models are shown with three different colors, with values of $\chi^2$ per degree of freedom (DOF) showing at top-right of the left panel to quantify the residuals.
     }
     \label{fig:example_mock_step}
\end{figure*}

\begin{figure*}
    \centering
    \includegraphics[width=1.0\textwidth]{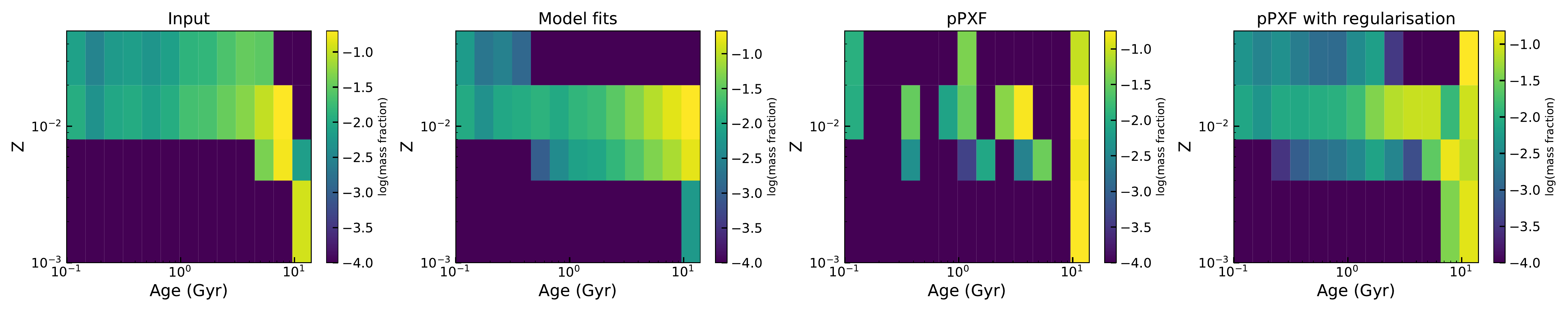}\\
     \caption{Comparison between our model fits with {\tt pPXF}. Plots show the mass fractions of SSPs in the metallicity-age space. From left to right, 
     the first panel plots the input model that is used to calculate the mock spectrum, corresponding to black lines in \autoref{fig:example_mock_step}; best-fit results recovered using model 1 (red lines in \autoref{fig:example_mock_step}) are shown in the second panel; the third panel comes from {\tt pPXF} fitting of the mock spectrum without regularization, while the last panel shows the results from regularized {\tt pPXF} fits.
     }
     \label{fig:example_mock_step_ppxf}
\end{figure*}

\begin{figure*}
    \centering
    \includegraphics[width=1.0\textwidth]{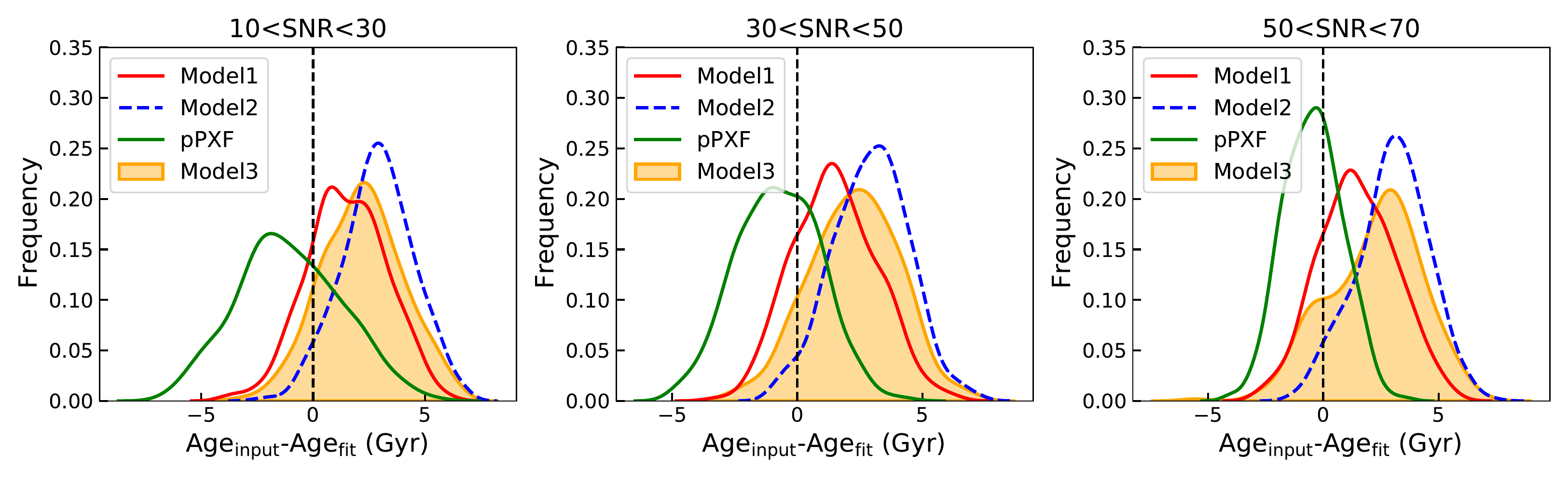}\\
    \includegraphics[width=1.0\textwidth]{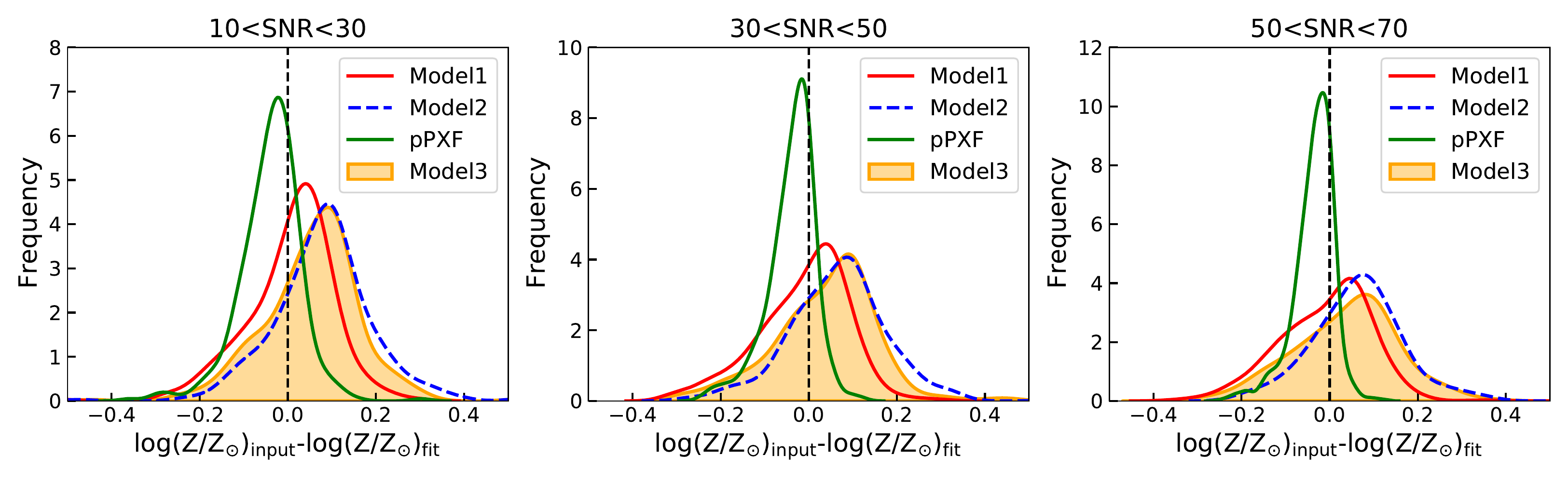}\\
     \caption{The distribution of the difference in the mean age (top) and stellar metallicity (bottom)
     between the best-fit model and input values. The three columns are results from fitting mock spectra with different levels of SNR, as indicated. Results obtained from the three models and results from {\tt pPXF} with regularisation are shown with different colors as indicated. Probability distributions are each normalized to 1.}
     \label{fig:mock_stats_step}
\end{figure*}

Of course, this is an idealized case, as a real galaxy may follow an evolutionary history that is not exactly matched by the assumptions of our model.  Therefore, as an extreme test, we have generated a further set of non-physically-motivated random histories in which we simply assume 
a random discrete evolution in the SFH and ChEH.  We adopt 7 time interval bins at look back time $0\to 0.5$, $0.5\to 1.0$, $1\to 2$, 
$2\to 4$, $4\to 6$, $6\to 10$ and $10 \to 14$ Gyr, and the SFH and ChEH are specified by the average SFRs and gas phase metallicities in each interval, respectively. For the SFH, we assume the average SFRs are randomly distributed between 0 and 1 in each time interval.
For the ChEH, the only constraint we place on the ChEH is that it increases over time, but is otherwise randomly selected.  As observations at different redshift have shown that the gas phase metallicities are becoming higher in galaxies at lower redshift \citep{Ly2016,Sanders2021}, this monotonically increasing ChEH should at least be a reasonable choice when trying to emulate the real evolution. To ensure this monotonicity, we first randomly generate the metallicity in each time interval between 0 and 0.05 (the highest metallicity available for our SSP templates), and then for each interval its metallicity will be set to the value in the nearest older bin if this random process gives it a lower metallicity. As in the previous test, the mock SFH and ChEH are combined with the  BC03 SSP templates to calculate mock spectra. Finally the spectra with Gaussian noise added are fitting with the three models.

\autoref{fig:example_mock_step} shows an example of fitting such a random mock spectrum. As might be expected for such a parametric fit, we are now unable to reproduce the input SFH and ChEH exactly.  However, the level of flexibility in our full model (model 1) means that it still does a respectable job of characterising the evolution of even such a completely random data set, and the quality of fit to the spectrum is significantly better than that of the two simplified versions of the model (see the $\chi^2$ per degree of freedom indicated in the plot). We also compare the fitting results of this spectrum with those obtained using {\tt pPXF}, which is shown in \autoref{fig:example_mock_step_ppxf}. Again as might be expected, there is a trade-off between the physically-motivated smoothness that the model imposes and a level of bias that constraining the fit in this way causes.  Nonetheless, the parametric fit does a respectable job of reproducing the general history of the galaxy even in this extreme unphysical case.

As before, we randomly generate and fit 1000 such mock spectra, and derive the differences between the input and best-fit results; the top and bottom panels of \autoref{fig:mock_stats_step} show the distributions of the differences in the mean ages and metallicities respectively.  As discussed, the imposition of a parametric model introduces a bias that reflects its physical motivation when compared to an unphysical random model, but it is clear that the full model 1 reduces that bias significantly compared to the simplified models, resulting in fits that, at least on average, reproduce the properties of the input data reasonably well. Again, results derived using {\tt pPXF} with regularization are shown in green.
In this case, as expected, results derived from {\tt pPXF} are better than all
the three models, owing to its greater flexibility.

We should also bear in mind that there are potentially other residual sources of uncertainty and bias that are embedded within this process. For example, we have assumed a simple constant uncertainty in the gas-phase metallicity and SFR as derived from emission lines, but there may well be systematic effects that still remain in such measurements, whereby the metallicity of the ionised phase is not entirely representative of the galaxy.  In addition, there are many different SSP templates that we could have used \citep[e.g.][]{BC03,Maraston2005,Vazdekis2010}.  Here, we have adopted the widely-used BC03 models, but a different choice would likely affect the results at some systematic level.  

Nonetheless, the approach adopted here, fitting a simple model that incorporates the main physical processes of star formation, inflow and outflow, seems to produce robust consistent results using data of a quality that can realistically be obtained.  As we will see in \autoref{sec:results}, there are a number of consistency checks that can be made with studies both in the local Universe and as a function of redshift, which allow us to test the validity of the results obtained, and hence, ultimately, the credibility of the model itself.

\section{Data}
\label{sec:data}
As an initial real-world application of this method, we have applied it to a sample of spiral galaxies selected from the SDSS-IV MaNGA survey. Such galaxies are likely to contain both emission and absorption lines in their spectra, so will test the effectiveness of an approach that can combine such information.  In addition, spiral galaxies are believed to have formed as well-defined single systems over most of their lifetimes, so will be less likely to have their evolutionary history confused by multiple mergers.  In this section we will give a brief introduction to the MaNGA survey and discuss the sample selection and data reduction process.

\subsection{MaNGA}

MaNGA (Mapping Nearby Galaxies at Apache Point Observatory) is one of the three core programmes of the fourth generation of SDSS (SDSS-IV, \citealt{Blanton2017}).
MaNGA has obtained spatially resolved, high quality spectra of more than 10,000 galaxies in the local Universe (redshift range $0.01<z<0.15$ \citealt{Yana2016,Wake2017}). Targets of MaNGA are selected from the NASA Sloan Atlas catalogue
\footnote{\label{foot:nsa}\url{ http://www.nsatlas.org/}} (NSA, \citealt{Blanton2005}). The whole sample covers the stellar mass range $5\times10^8 {\rm M}_{\odot}h^{-2} \leq M_*\leq 3 \times 10^{11} 
{\rm M}_{\odot}h^{-2}$ with a roughly flat number density distribution \citep{Wake2017}. Targets are covered by MaNGA out to a radius of either $1.5 R_{\rm e}$ or $2.5 R_{\rm e}$ ($R_{\rm e}$ being the effective radius) for the “Primary” and “Secondary” samples, respectively \citep{Law2015}.  Light from the galaxies is collected by the Sloan 2.5m telescope \citep{Gunn2006} and sent to the two dual-channel BOSS spectrographs
\citep{smee2013} to produce spectra covering $3600-10300${\AA} in wavelength, with spectral resolution $R\sim2000$ \citep{Drory2015}. Readers are referred to \cite{Yanb2016} for the spectrophotometry calibration details of MaNGA, while the initial performance is presented in detail in \cite{Yana2016}.

The raw MaNGA data was reduced and calibrated by the Data Reduction Pipeline (DRP; \citealt{Law2016}). The DRP produces data cubes containing science-ready spectra with a relative flux calibration better than $5\%$ in more than $80\%$ of the wavelength range \citep{Yana2016}. In addition, MaNGA provides stellar kinematic measurements, emission-line properties, and spectral indices for each spaxel obtained through the MaNGA Data Analysis Pipeline \citep[DAP;][]{Westfall2019,Belfiore2019}. 

\subsection{Sample selection and data reduction}
\label{sec:data_re}

In this work, we select a sample from the 11th MaNGA Product Launch (MPL11) data. MPL11 includes all the samples observed by MaNGA, which contains 10,010 high-quality, unique galaxies. It has been  released with the SDSS-IV Data Release 17 (DR17, \citealt{SDSSDR17}), which is also the final release of MaNGA data . To select a suitable sample of spiral galaxies, we make use of a value-added catalogue, MaNGA Visual Morphologies from SDSS and DESI images. This catalogue classified all the MPL11 galaxies morphologically based on inspection of their mosaic images generated from a combination of the SDSS and the Dark Energy Legacy Survey (DESI) images. Readers are referred to \cite{Lacerna2020} for the introduction to this value-added catalog (VAC) for the MPL-7/DR15 version; the final version has yet to be publicly released.  In this VAC, each galaxy is assigned a $T$-type number to quantify its morphology, with $T$-type$\leq$0 for early-type galaxies, $T$-type$>$0 for late-type galaxies.  We thus select spiral galaxies through requiring $T$-type$>$0. In addition, to minimise inclination and dust attenuation effects, we require the sample galaxies to be reasonably face-on, so that galaxies with axis ratio $\frac{b}{a}<0.5$ are excluded. Finally, to ensure that the spectral absorption-line and emission-line fits are reasonably well constrained, we require the spectra co-added within 1$R_{\rm e}$ of the galaxy to have a signal-to-noise ratio SNR$>30$, and at least 10 pixels within 1$R_{\rm e}$ to have SNR$>5$ in the four emission lines \hbox{[O\,{\sc iii}]}$\lambda$5007, \hbox{[N\,{\sc ii}]}$\lambda$6584, H${\alpha}$, and  H${\beta}$. This selection process generated a final sample of 2560 galaxies.
  
In the raw data cubes provided by the DRP, original spectra have typical $r$-band SNRs of $4-8$ {\AA}$^{-1}$ at the outskirts of galaxies \citep{Law2016}, which is too low for detailed analysis of metallicity. In this work, we use all the spaxels within 1$R_{\rm e}$, and stack their MaNGA spectra to achieve sufficient SNR for the subsequent analysis, allowing us to calculate global integrated properties for each galaxy with some confidence. The stacking procedure is similar to \cite{Zhou2019}, and we refer the reader to the paper for full details. During the stacking,  the spectra are shifted to the rest-frame, and in what follows all wavelengths mentioned are rest-frame values unless explicitly stated to be otherwise. In addition, we correct for the effects of co-variance between spaxels using the correction term given by \cite{Westfall2019}. After this correction, the final SNR (averaged over all wavelengths for a given spectrum) of our stacked spectra is typically around 70 per {\AA}, which, as we have seen, is well suited to deriving reliable estimates of galaxies SFHs and ChEHs.

We also make use of the current SFR estimated from the H${\alpha}$ fluxes as an additional constraint. To this end, we used H${\alpha}$ flux measurements of every individual pixels from the DAP.
Those fluxes are corrected for dust attenuation using the Balmer decrement, adopting the Calzetti extinction law \citep{Calzetti2000} and assuming a intrinsic H${\alpha}$/H${\beta}$ ratio of 2.87 \citep{Osterbrock2006}. After the dust correction,  fluxes from pixels within 1$R_{\rm e}$ are combined to derive a total H${\alpha}$ flux, which are converted to the SFR using the calibration of \cite{Murphy2011} and assuming a Chabrier \citep{Chabrier2003} IMF:
\begin{equation}
\label{eq:sfr}
\rm SFR(M_{\odot}yr^{-1})=5.37\times10^{-42}L(H\alpha).
\end{equation}
Given the thousands of individual pixels available for a MaNGA galaxy, we found that the statistical uncertainties for this SFR measurement is typically less than 2\%. Note that there might be other sources of photoionisation that would produce H${\alpha}$ emission. One could account for such extra emission by treating the SFR constraint as an upper limit in a modified version of \autoref{likelyhood}.  However, the current sample galaxies have been selected to be at least reasonably star-forming, so such contamination from other sources should be relatively small.  We therefore prefer to keep the likelihood function simple, and \autoref{likelyhood} is applied throughout this proof-of-concept analysis.

Similarly, to derive the current gas phase metallicity of the sample galaxies, we make use of the emission line measurements provided by the DAP. This pipeline fits simultaneously the continuum and emission lines for every MaNGA spectrum, and the emission line fluxes are provided after subtracting the stellar continuum model to correct for underlying stellar absorption lines [see \cite{Belfiore2019} for more details]. There are a range of available indicators and calibrations to characterise the gas phase metallicity \citep[e.g.][]{PP04,Maiolino2008}. For convenience in comparing with observations of MZR at high redshift \citep[e.g.][]{Sanders2018,Topping2021}, we chose to adopt the $O3N2$ index as a metallicity indicator. However, our tests have shown that the major results of this work are not strongly dependent on the specific indicator adopted. The $O3N2$ index is defined as 
\begin{equation}
 O3N2 \equiv
\rm \log \frac{\hbox{[O\,{\sc iii}]} \lambda 5007/{ H}{\beta}}{\hbox{[N\,{\sc ii}]} \lambda 6584/H{\alpha}}.   
\end{equation}
We use the calibration of \cite{PP04} to convert the $O3N2$ index into oxygen abundance,
\begin{equation}
12+\log({\rm O/H})=8.73-0.32*O3N2.
\end{equation}
From this calibration, we derived the oxygen abundance of all spaxels within 1$R_{\rm e}$ of the galaxy, and simply use the median value as the current gas phase metallicity of the galaxy. Again, the large number of individual pixels in MaNGA galaxies allows a reliable estimate of the median gas phase metallicity, with statistical errors typically less than 2\%.

\begin{figure*}
    \centering
    \includegraphics[width=1.0\textwidth]{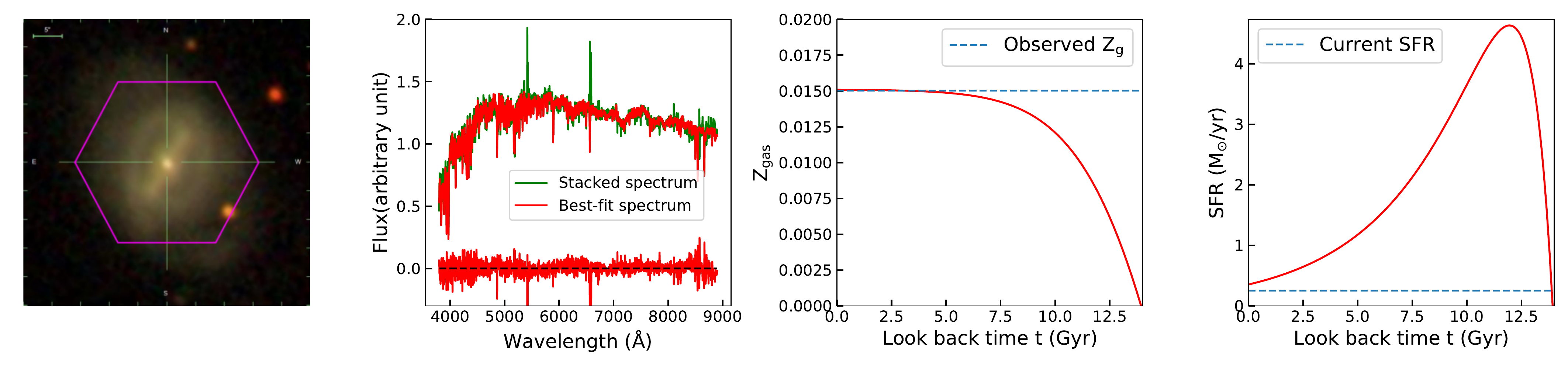}\\
    \includegraphics[width=1.0\textwidth]{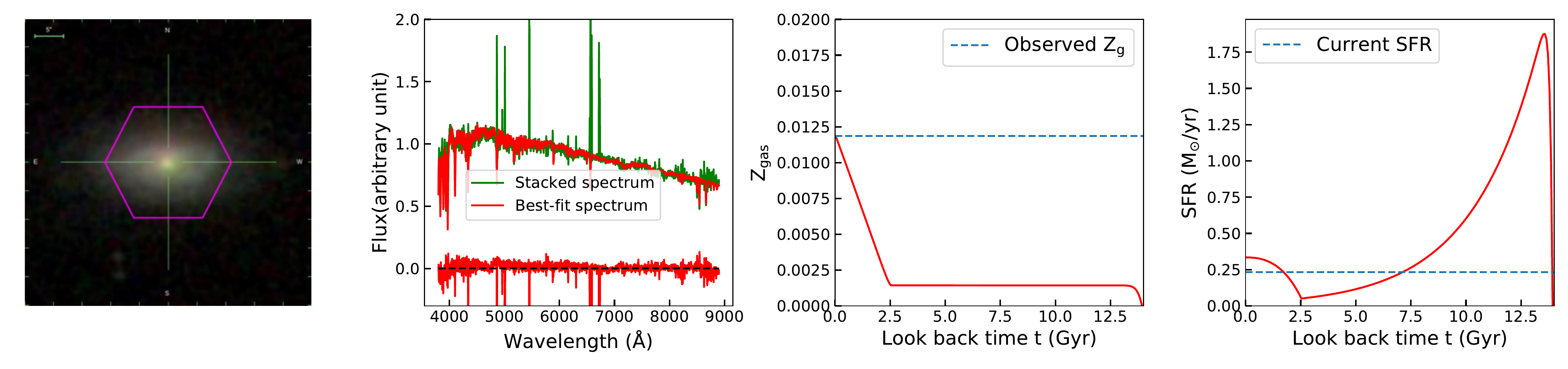}
     \caption{Two typical galaxies in our sample. From left to right, the first panels show the optical images of the two galaxies, with the MaNGA footprint shown in magenta. The second panels compare our best-fit spectrum (red) with the observed spectrum (stacked within 1$R_{\rm e}$ of the galaxy, green), with residuals shown at the bottom. The third panels are the ChEHs calculated from the best-fit parameters, with blue dash lines indicating the observed current gas phase metallicity from the O3N2 estimator. The last panels show the normalized SFHs of the two galaxies, with blue dash lines indicating the current SFR of the galaxy estimated from its H${\alpha}$ fluxes.
     }
     \label{fig:example}
\end{figure*}

To facilitate a direct comparison between the metallicity calculated by the chemical evolution model (see \autoref{sec:method}), the gas-phase metallicity calculated from the $O3N2$ index, and the stellar metallicity from SPS methods, we normalized all the metallicities using the solar oxygen abundance. \cite{Asplund2009} recommend a solar metallicity of 0.014 and oxygen abundance of $12+\log{\rm O/H})=8.69$.
However, in this work we use the \citet{BC03} SSP models calculated with the ‘Padova1994’ stellar evolutionary tracks, in which an older calibration with solar metallicity of 0.02 and oxygen abundance of $12+\log({\rm O/H})=8.83$ \citep{Anders1989} is applied. To be consistent with the SSP template settings, we adopt this older value in what follows, and appropriate caution is urged when comparing with other calibrations.

\section{Results}
\label{sec:results}

We are now in a position to apply the semi-analytic spectral fitting method to this well-defined sample of spiral galaxies.  Having derived the ChEHs and SFHs of spiral galaxies, and how they vary with properties like their masses, we can compare them to other analyses both in the nearby Universe and as a function of redshift, as an independent test of the plausibility of the model we have fitted.  Once this credibility is established, we will be able to move on to interpret the parameters of the model themselves.

Two examples, illustrating the range of properties that this very general model can reproduce, are presented in \autoref{fig:example}. For the galaxy in the top panels, it is seen that this system experiences only one major star formation event, and its star formation rate steadily decreases with time, while its gas-phase metallicity gradually increases to the current value. The evolution is similar to the model presented in \autoref{fig:model_example_outflow}, which indicates that the evolution of this galaxy is regulated by long-lasting gas inflow and moderate outflow. In contrast, for the galaxy in the bottom panels, two major star formation events are seen. The first one happened very early due to strong gas infall, while the second occurred around $3\,{\rm Gyrs}$ ago. Its metallicity quickly increased to around 10\% solar and remained constant for 8 Gyrs, followed by a further rapid increasing in the past $3\,{\rm Gyrs}$. According to \autoref{fig:model_example_outflow_turnoff}, this evolution indicates that the galaxy experienced a strong outflow in the early Universe which suppressed further enrichment of its metal content. When that outflow ended around $3\,{\rm Gyrs}$ ago,  the galaxy rejuvenated and entered a secondary star formation and chemical enrichment phase. In what follows, we will explore how typical such evolutionary histories are, how they vary systematically between galaxies, and how they fit with other measured properties.

\subsection{Evolution of mass and metals: downsizing formation}

\subsubsection{The star formation history}
\begin{figure}
    \centering
    \includegraphics[width=0.4\textwidth]{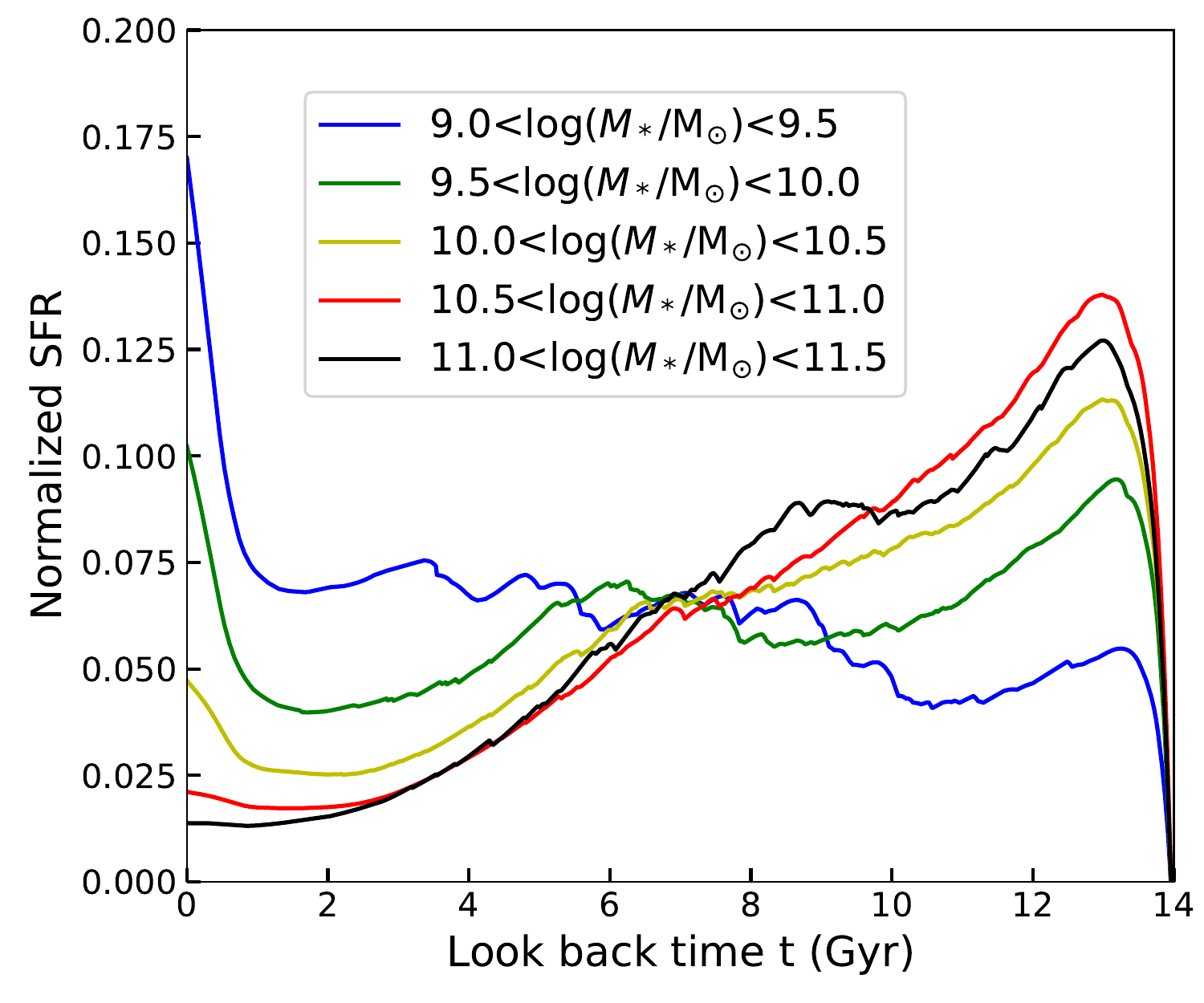}\\
     \caption{The star formation histories of our sample galaxies. Lines are the mean of the SFHs in five stellar mass bins, as labelled. The SFHs are normalized to form 1M$_{\odot}$ in total.}
     \label{fig:sfh}
\end{figure}

We first focus on the derived SFHs in our sample. \autoref{fig:sfh} shows the average SFHs of the sample galaxies in five stellar mass bins. The derived SFHs of the galaxies show a clear dependence on their stellar masses. The formation for galaxies with $M_*>10^{10.5} {\rm M}_{\odot}$ is quite simple, with the SFHs peaking around $12\,{\rm Gyr}$ ago and then decreasing exponentially with time. In contrast, the formation of less massive galaxies becomes more complex. Those galaxies also experienced significant star formation around $10\,{\rm Gyr}$ ago, but they are not quenched entirely after the initial surge. Instead, many galaxies experience a secondary star formation event, and their SFR keeps increasing in the most recent $4\,{\rm Gyr}$. This mass dependence of galaxies' SFHs is consistent with many previous investigations \citep{Panter2003,Kauffmann2003,Heavens2004,Panter2007,Fontanot2009,Peng2010,Muzzin2013}, illustrating the downsizing formation scenario in which more massive galaxies form their stellar masses earlier, while less massive galaxies have relatively younger stellar populations. However, despite the blue colors seen in the lower mass galaxies ($M_*<10^{9.5} {\rm M}_{\odot}$) in the sample, our results reveal that those galaxies formed a significant fraction of their stellar masses more than $8\,{\rm Gyrs}$ ago, which is consistent with our previous results based on direct modelling of the SFH of low mass galaxies \citep{Zhou2020}, and is also in line with resolved observations for some local dwarf systems \citep{Weisz2011}.

\begin{figure}
    \centering
    \includegraphics[width=0.4\textwidth]{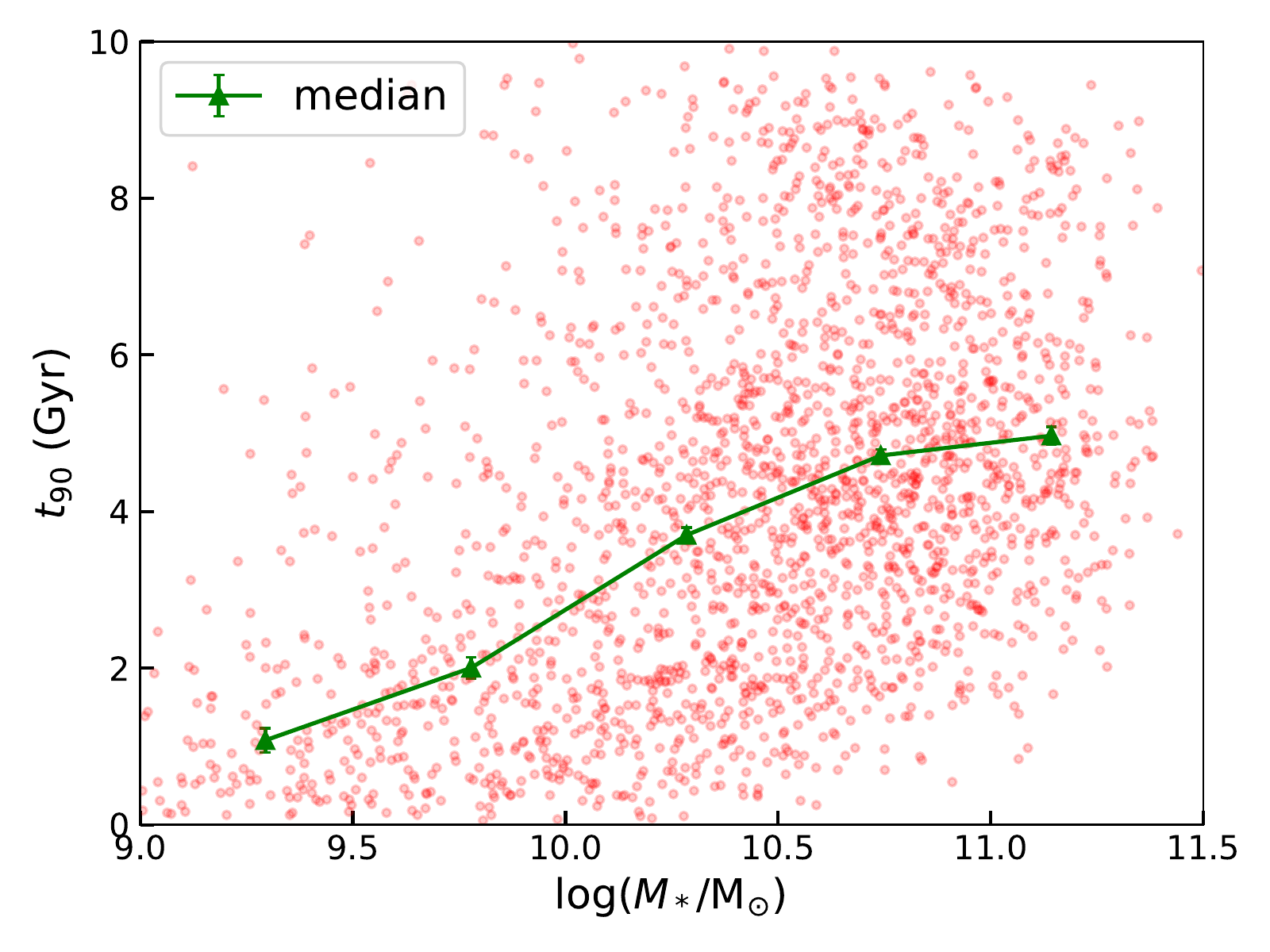}\\
    \includegraphics[width=0.4\textwidth]{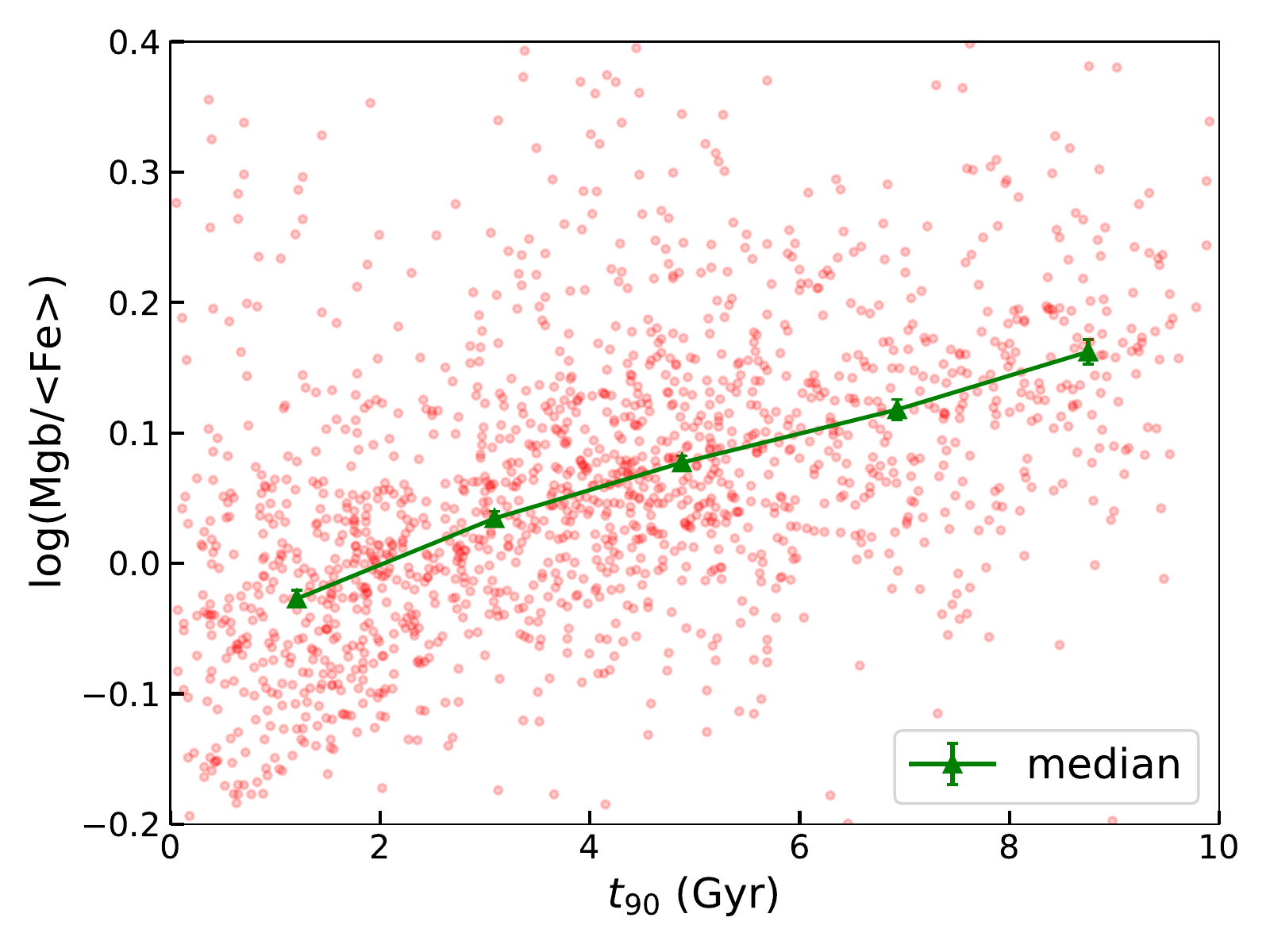}
     \caption{Correlations between the 90\% stellar mass formation time ($t_{\rm 90}$) and the current stellar mass ($M_*$, top panel) and $\alpha$-element abundance characterised by the $\rm Mgb/\langle Fe\rangle$ index (bottom panel) of our sample galaxies.  In each panel, green triangles linked by lines are medians in different bins, with error bar showing the error  estimated  from  the  jackknife resampling method.}
     \label{fig:t_half}
\end{figure}

To better quantify the scale of downsizing in a robust manner, \autoref{fig:t_half} shows the time it took the galaxies to form 90\% of their stellar masses (called $t_{\rm 90}$ hereafter) as a function of their stellar masses. Typically, galaxies with  $M_*>10^{10.5} {\rm M}_{\odot}$ formed most of their stellar mass more than $4 - 6\,{\rm Gyr}$ ago, while the least massive ones ($M_*<10^{9.5} {\rm M}_{\odot}$) completed the formation of 90\% of their current mass only about $1\,{\rm Gyr}$ ago.

As a further check on the reality of the compressed timescale for the formation of massive galaxies,
we plot in \autoref{fig:t_half} the correlation between $t_{\rm 90}$ and $\rm Mgb/\langle Fe\rangle \equiv Mgb/(0.5*Fe5270+0.5*Fe5335)$ in the sample, in which the Mgb, Fe5270 and Fe5335 values are obtained from the MaNGA DAP. $\rm Mgb/\langle Fe\rangle$ is a proxy 
of the $\alpha$/Fe ratio, which is often used as an indicator of the timescale of the star formation process \citep[e.g.][]{Worthey1994,Thomas2005,Zheng_etal2019}.The Mgb index traces the abundance of $\alpha$-elements that are produced in core-collapse supernovae, while the Fe5270 and Fe5335 indices are indicators of the iron abundance that are  generated both in core-collapse and in type Ia supernovae \citep{Nomoto1984,Thielemann1996}. Since the explosion of of type II (core collapse) supernova happens very shortly after star-formation, while low-mass stars need a longer time to evolve into the progenitors of type Ia supernovae, a high relative abundance of $\alpha$-elements indicates that stars form on a short timescale so that the ISM has not been polluted by type Ia supernova explosions \citep{Thomas2005}.  It is apparent from \autoref{fig:t_half} that the derived $t_{\rm 90}$ is strongly correlated with $\rm Mgb/\langle Fe\rangle$ of the galaxy, such that galaxies with lower $\rm Mgb/\langle Fe\rangle$ have shorter $t_{\rm 90}$. As indicated by the top panel, galaxies with $t_{\rm 90}\sim 1\,{\rm Gyr}$ are the least massive galaxies. And according to \autoref{fig:sfh}, those galaxies have experienced multiple star formation events and have an extended SFH, which is also reflected in the lower $\rm Mgb/\langle Fe\rangle$. In contrast, as has been known for many years \citep[e.g.][]{Worthey1992}, massive galaxies are enhanced in their $\alpha$-element abundance. \autoref{fig:t_half} shows that this enhancement is tightly correlated with their shorter formation times. The consistency of the picture emerging suggests that the derived SFHs characterise the galaxies' mass growth well, with $t_{\rm 90}$ providing a good indicator of their star-formation timescales.

In summary, the derived SFHs are in good agreement with the known result that massive galaxies formed earlier than low mass ones. In this semi-analytic spectral fitting, it is notable that this result is a byproduct of a physical model that follows the evolution of gas content in galaxies.  The fact that this simple model can reproduce such population properties lends weight to its use in beginning to understand the relative importance of the different physical processes that it incorporates.

\subsubsection{The chemical evolution}
 \begin{figure}
    \centering
    \includegraphics[width=0.4\textwidth]{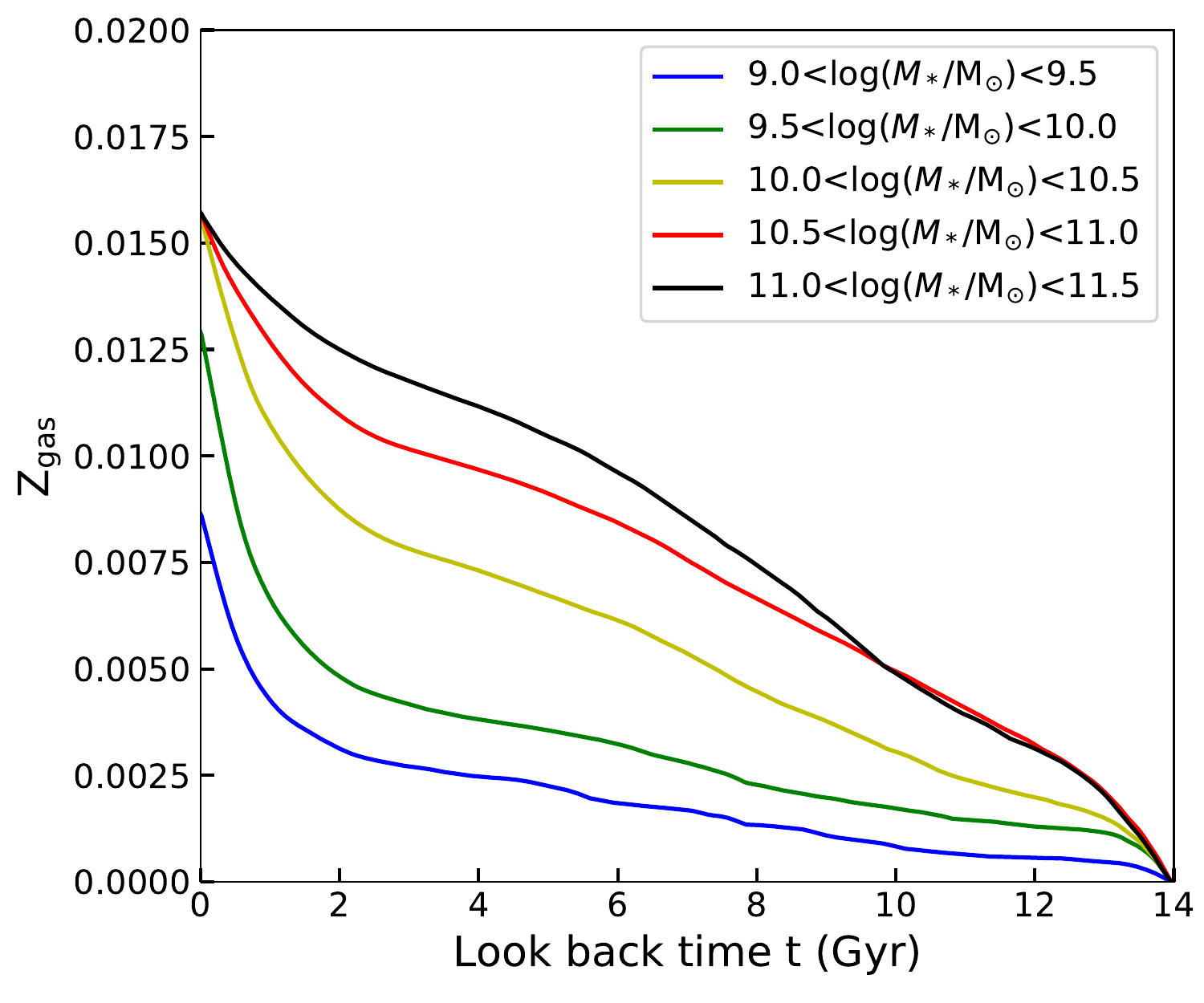}
     \caption{The chemical evolution histories of our sample galaxies.  Lines are mean ChEHs in five stellar mass bins as labelled. }
     \label{fig:ChEH}
\end{figure}

We now turn to the chemical evolution of the sample galaxies. \autoref{fig:ChEH} shows their mean ChEHs in five stellar mass bins. As with the SFH, the ChEH of a galaxy is strongly correlated with stellar mass. Massive galaxies ($M_*\gtrsim10^{10.5} {\rm M}_{\odot}$) show a gradual accumulation of metals, connected to their ongoing star formation since early times.  This kind of evolution is similar to the model presented in \autoref{fig:model_example_outflow}, which is driven by rapid initial gas infall and weak outflow.  By contrast, low mass galaxies experience a very different chemical evolution path. Although star formation was also initiated quite early in those galaxies, their metallicities remain at a relatively low value. Around $4\,{\rm Gyr}$ ago, associated with a second star formation event (see \autoref{fig:sfh}), metallicities in these galaxies increase relatively rapidly and reach the current gas phase metallicity observed by MaNGA. As shown in \autoref{fig:model_example_outflow_turnoff}, this two-phase evolution is driven by the time dependence of their outflows. Such galaxies experience strong outflows in the early Universe, which suppress their early metallicity enrichment processes. These outflows turn off in more recent times, triggering the secondary star formation event associated with a fast chemical enrichment process. Although this late-time enrichment increases the metallicities of low-mass systems, over all it is not as effective as the steady enrichment seen in higher-mass systems, so that the present-day metallicity of their gas phase remains lower, as seen in various previous studies \citep[e.g.][]{Gallazzi2005,Panter2007,Thomas2010}. Notably, \cite{CampsFarina2021CALIFA} and \cite{CampsFarina2022} are also currently using the fossil record methods to derive the ChEH in CALIFA and MaNGA galaxies. Despite the many differences in methods, data, and templates between their work and this approach, they also detect indications of a rapid increase in metallicity over the last $\sim 4\,{\rm Gyr}$.  This consistency offers confidence in the robustness of both methods, and the plausibility of the physical interpretation of this finding.

\subsection{Comparing the results to other analyses}
Through this semi-analytic approach, we can start to infer the essential physical processes that drive the chemical evolution of galaxies.  However, one concern is that the model is too simple to capture the true physics involved, and is not sufficiently general to reproduce this evolution. One way we can test the validity of this concern is by comparing the results we obtain to those previously derived without such physical constraints, both in the nearby Universe and at higher redshifts.
 
\subsubsection{The cumulative metallicity distribution function}
\label{sec:cmdf}

\begin{figure}
    \centering
    \includegraphics[width=0.4\textwidth]{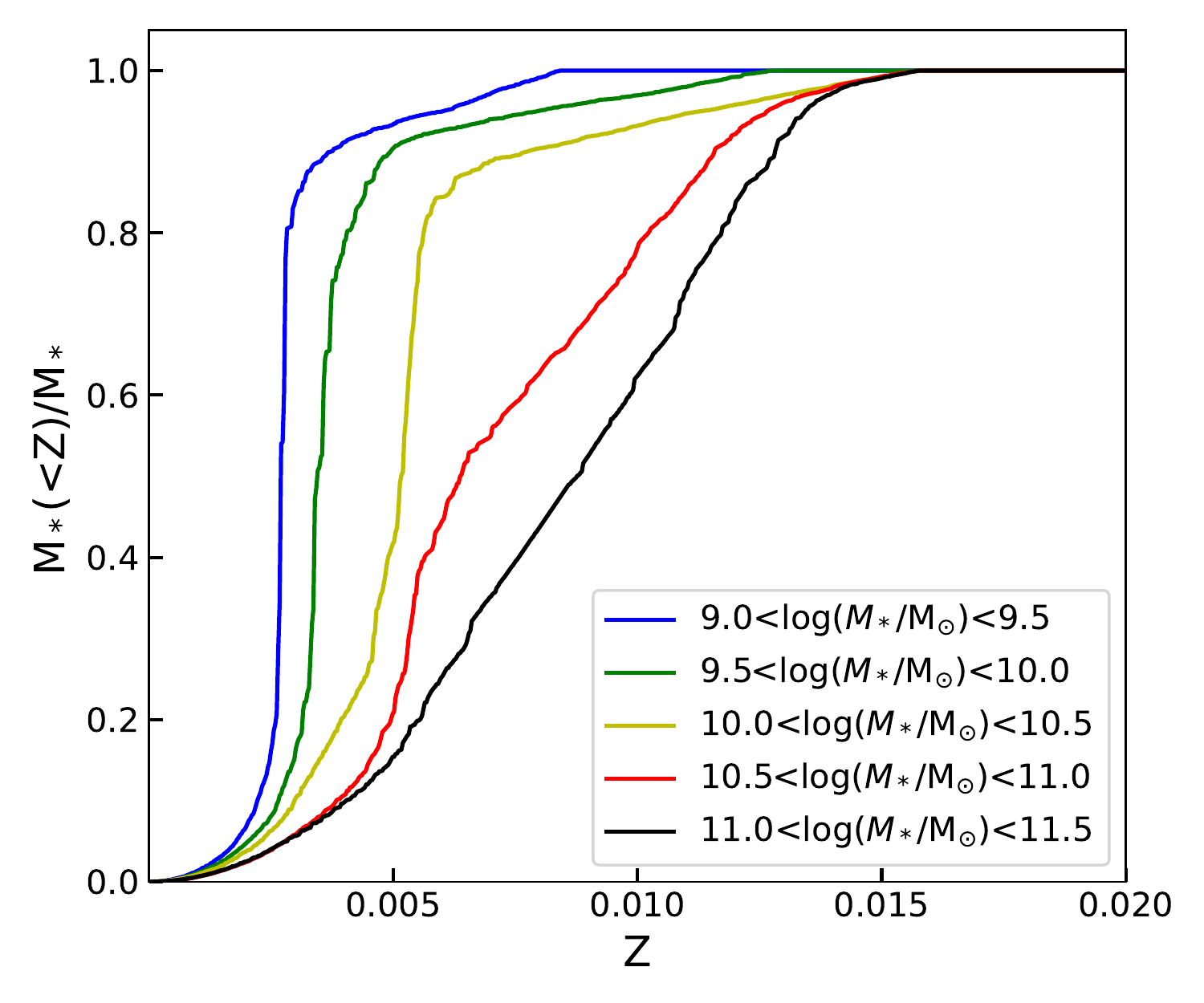}
     \caption{The Cumulative metallicity distribution function of the sample galaxies. Lines are mean CMDFs in five stellar mass bins as labelled.}
     \label{fig:CMDF}
\end{figure}

Perhaps the simplest chemical property that can be derived for a galaxy's stellar population is its cumulative metallicity distribution, i.e. the total mass of stars with metallicity less that Z, $M_*(<Z)$. This simple function loses much of the detailed information in the ChEH, but it provides a robust measure that still tells us something about a galaxy's history.  For example, the relative paucity of metal-poor stars in the Milky Way has long been known to be inconsistent with a simple closed-box model, implying the presence of gas inflows and outflows.  Since this effect was first discovered in local G dwarf stars, it is generally known as "the G-dwarf problem" \citep{vandenBergh1962}.  Recently, \citet{Greener2021} have investigated this phenomenon in other galaxies using MaNGA data, by estimating $M_*(<Z)$ using conventional spectral fitting.  They found that massive spiral galaxies generically display this effect in that $M_*(<Z)$ rises only slowly with $Z$ for small $Z$, but low-mass galaxies do not.  In \autoref{fig:CMDF}, we produce the equivalent plot from our semi-analytic spectral fitting.  It shows both qualitative and quantitative agreement with the work of \citet{Greener2021}, confirming that the model we are using has sufficient flexibility to reproduce such observables.  The advantage we have is that we can now begin to understand what the main physical processes might be that cause galaxies to deviate from the simple closed-box picture, which we will discuss in \autoref{sec:interp}.

\subsubsection{The mass--metallicity relation at different redshifts}

\begin{figure}
    \centering
    \includegraphics[width=0.4\textwidth]{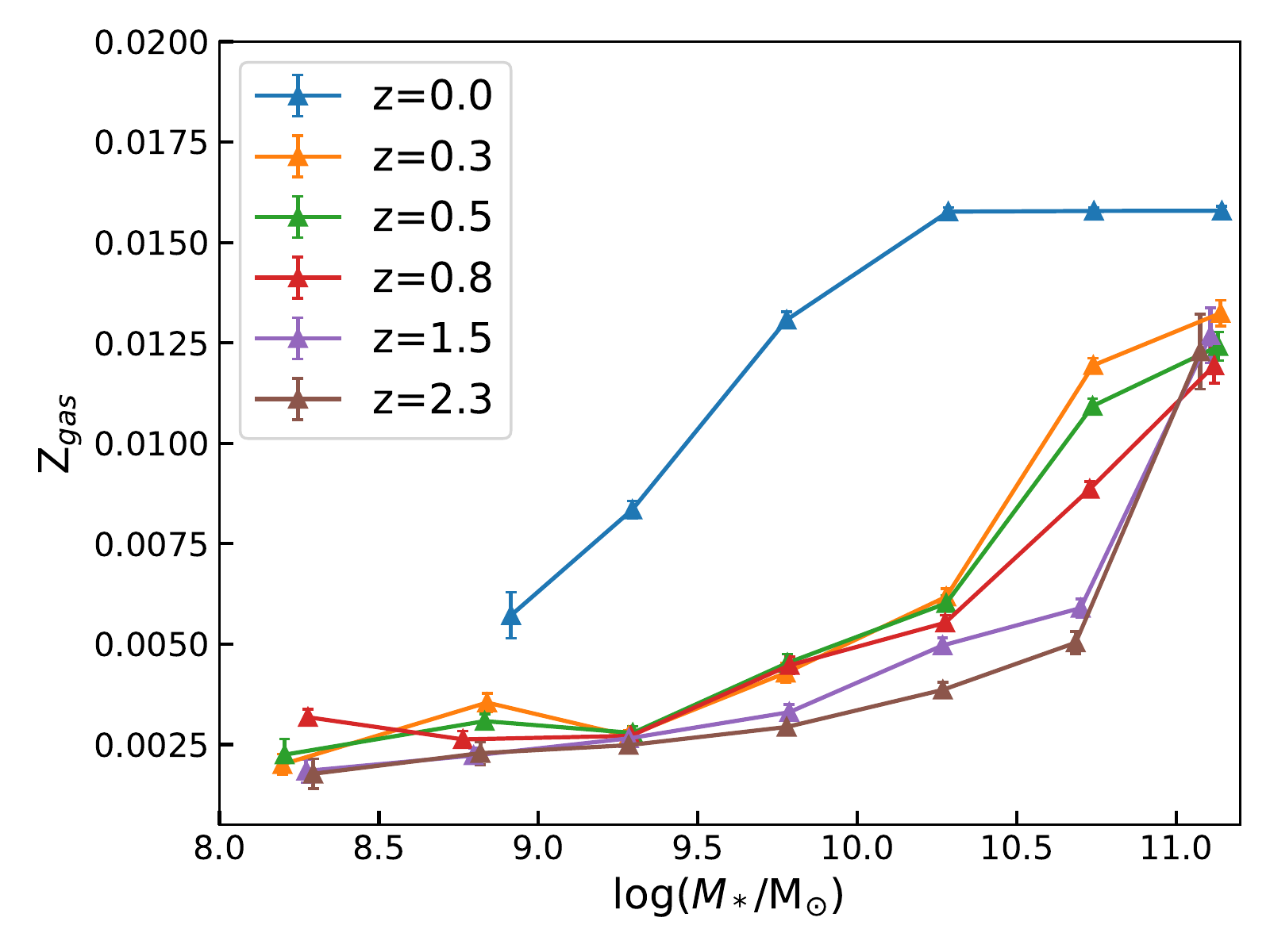}
     \caption{The redshift evolution of the mass--metallicity relation predicted by the model fits to the sample galaxies. Lines are the median values for galaxies in seven stellar mass bins, with statistical error estimated using the jackknife resampling method.}
     \label{fig:mzr_evolution}
\end{figure}

This semi-analytic approach makes specific predictions as to the gas metallicity at any redshift, which can be compared directly to observations. Unlike metallicity evolution constructed from the more conventional non-parametric approaches, this full evolutionary model directly predicts both the total accumulated stellar mass and the gas-phase metallicity at any redshift, and these quantities are physically linked throughout the evolution processes. This internal relation places additional constraints on the MZR.  We can therefore construct the evolution with redshift of MZR of the sample galaxies, and the results are presented in \autoref{fig:mzr_evolution}.

\begin{figure*}
    \centering
    \includegraphics[width=0.3\textwidth]{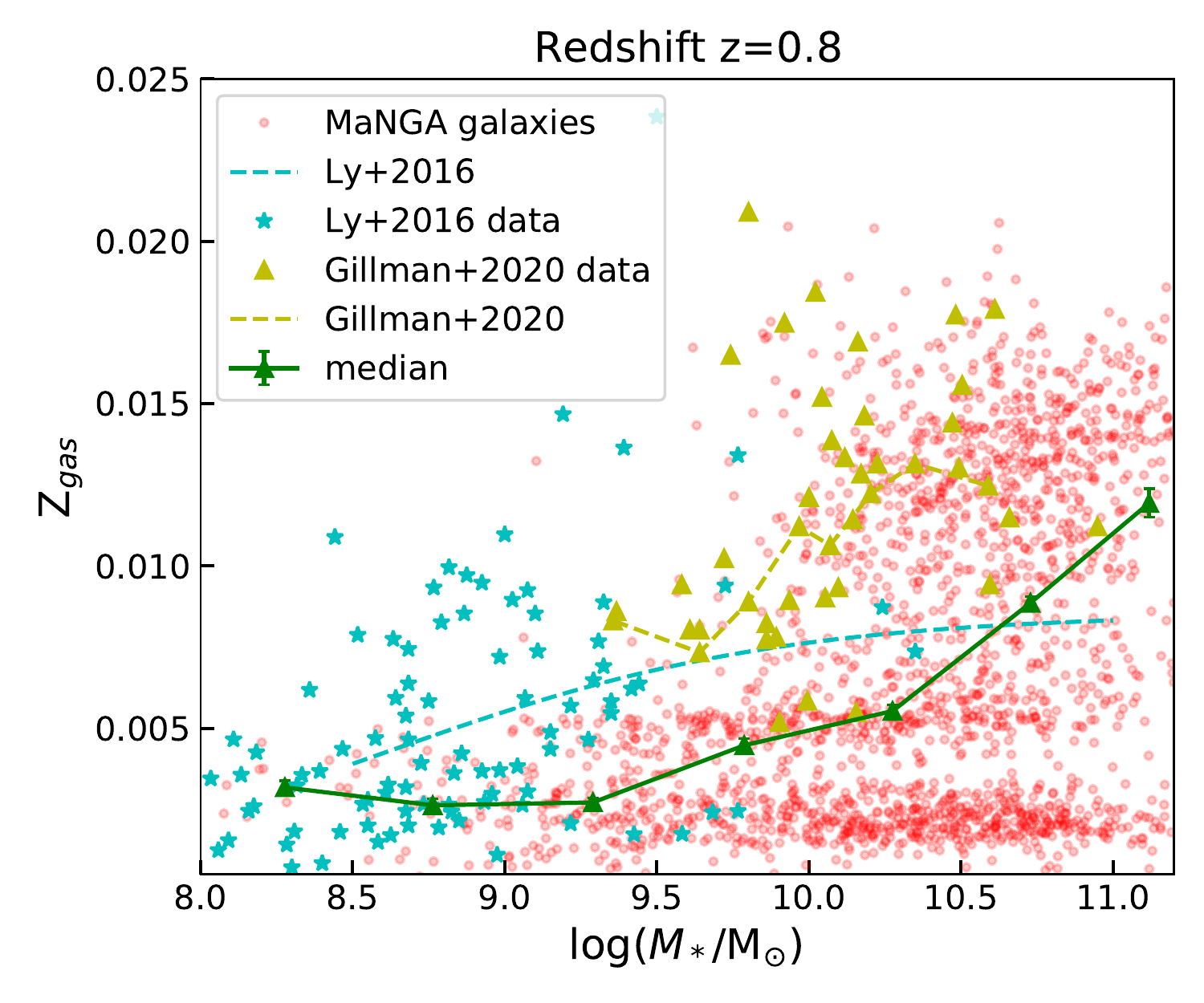}
    \includegraphics[width=0.3\textwidth]{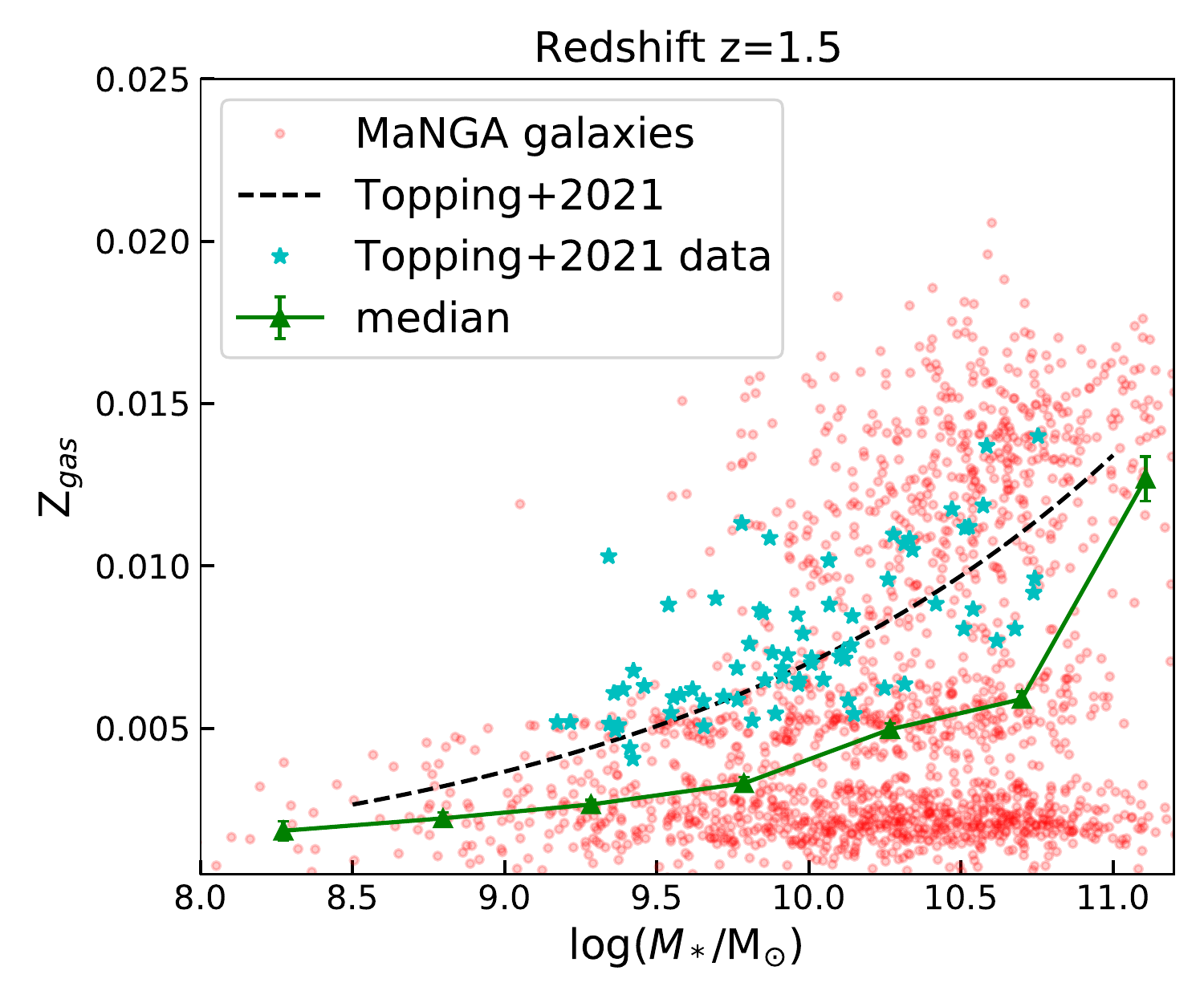}
    \includegraphics[width=0.3\textwidth]{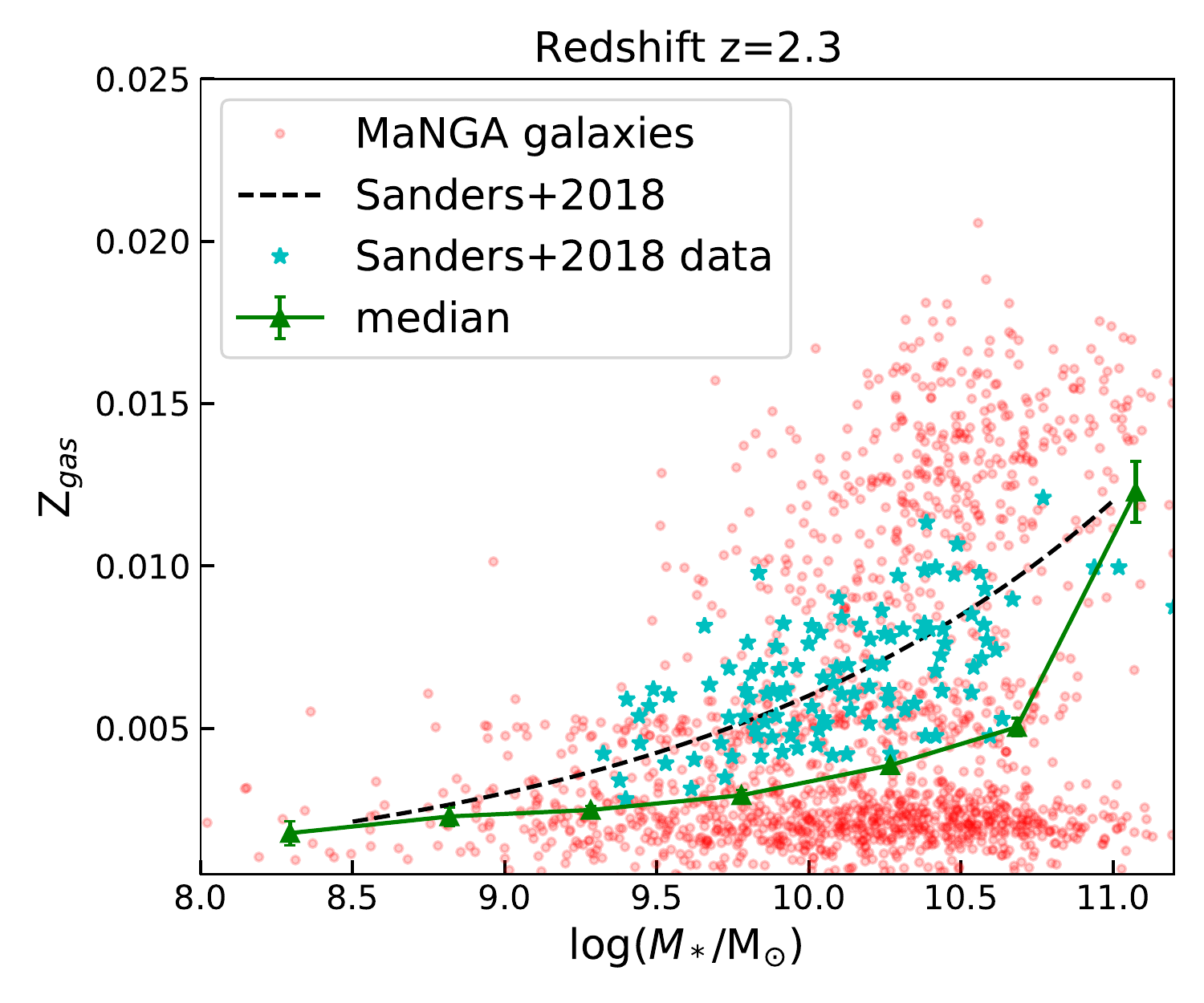}\\
     \caption{
     The MZR predicted from the best-fit SFHs and ChEHs for the sample galaxies at redshift $z=0.8$ (left), 1.5 (middle) and 2.3 (right). In each panel, red dots are our sample galaxies, green solid line indicates the median values in 7 stellar mass bins, with statistical error estimated from the jackknife resampling method. In the left panel, we plots the observation from \citet{Ly2016} and \citet{Gillman2021} in cyan and yellow respectively , for reference. Similarly, observations from \citet{Sanders2018} and \citet{Topping2021} are shown in the middle and right panels, respectively.
     }
     \label{fig:MZR_compare}
\end{figure*}

Perhaps the most striking feature of this plot is that although metallicity evolution is apparent at all redshifts, it has accelerated in the recent past, between redshift 0.3 and the present day.  This surprisingly recent change has also been seen in direct observations.  For example, \cite{Ly2016} found that the MZR for their sample galaxies in the redshift range $0.3<z<0.5$ is very different from those obtained at $z<0.3$, while only small differences are seen between samples with $0.3<z<0.5$ and $0.5<z<1.0$. MZRs obtained at higher redshifts \cite[e.g.][]{Sanders2018,Topping2021} also have similar shape as those obtained from intermediate redshifts.

In addition, the redshift evolution presented in \autoref{fig:mzr_evolution} has a significant mass-dependence. The MZR of massive galaxies at $z\sim0.3$ is only slightly ($\sim$0.0025 for galaxies of $10^{11.0}M_{\odot}$) lower than the current value, while larger differences are seen for less massive galaxies ($\sim$0.01 for galaxies of $10^{10}M_{\odot}$). At redshifts higher than 0.3, the evolution of the MZR mainly happens at the high mass end ($>10^{10.0}M_{\odot}$), with the shape becoming shallower with time, indicating some kind of gradual saturation in metallicity. This mass dependence of the evolution of the MZR can be physically understood in this semi-analytic framework from the mass dependence of the SFHs and ChEHs in \autoref{fig:sfh} and \autoref{fig:ChEH}. The result is also consistent with \cite{CampsFarina2021CALIFA}, who found that the evolution of the stellar MZR for galaxies with stellar mass $M_*<10^{10.0}M_{\odot}$ mainly happened within the latest $5\,{\rm Gyr}$, while for high mass galaxies the evolution happened between $5\,{\rm Gyr}$ and $10\,{\rm Gyr}$ ago.

The changing shape of the MZR resulting from such differential evolution is also consistent with previous direct observations.  The zero-redshift form, in which metallicity initially rises rapidly with mass and then flattens above $\sim $10$^{10.5}{\rm M}_{\odot}$, has been observed in many previous studies \citep[e.g.][]{Tremonti2004,Kewley2008,Andrews2013}.  The rather different shape at higher redshift, which rises more steeply at higher masses, has also been directly observed \citep{Gillman2021, Topping2021, Sanders2018}.  

To attempt to make a more quantitative comparison to the literature, \autoref{fig:MZR_compare} shows all the sample galaxies in the MZR plane at redshifts matched to various samples from the literature.  The scatter in the individual points from our sample is large, presumably reflecting both the uncertainty in the measurements and the intrinsic variance in evolution followed by individual galaxies, but with a data set of this size the mean relations are well defined. One might also notice that there seems to be a subsample of galaxies that has quite low metallicity and show no MZR trend. Investigating further, we find that these galaxies are generally at the gap between their two star formation episodes. Their metallicity enrichment has been suppressed by a strong outflow, and they have not yet experienced the subsequent enrichment triggered by the outflow turn-off. As they are not significantly star-forming at the given redshift, such galaxies would likely not show in direct high-redshift observations and therefore probably do not contribute to the conventionally-observed MZR at that redshift, making a detailed comparison difficult. Moreover, a further challenge in comparing to higher redshift data is apparent in the left panel, where two different samples produce very different mean relations at high mass.  We also have no way of knowing how closely any of these objects correspond to the high-redshift analogues of our sample, as they were selected in completely different ways, and might also expect to be biased towards systems with unusually strong emission lines, as these are used to determine their metallicities. Nonetheless, it is reassuring that the scatter of high-redshift observations are found largely within the region that the semi-analytic model predicts that galaxies in our sample lay at the time, and the mean relations, although somewhat offset in value due to the different selection criteria, are similar in shape.

\subsubsection{Gas-to-stellar mass ratios}
\label{ssec:gas}
One other quantity that the semi-analytic model predicts is the total gas mass of each galaxy at the present day, which, of course, can be measured completely independently by direct observation as a further test of the model fit's credibility. To this end, we cross-matched our sample with galaxies in the HI-MaNGA survey for which 21cm observations have been made. We only include galaxies with good estimates of the total atomic gas mass from this survey, which yields a subsample of 537 galaxies. In principle we should also add the molecular and ionised components, but the total mass is typically dominated by this atomic component  \citep{Catinella2018}. The model only predicts the gas mass in the region that it is being fitted to, within $R_e$, so we also make a simple correction to this mass to predict the total mass of the system, $M_{\rm g,tot} \approx 10M_{\rm g, Re}$ \citep{Leroy2008}.

\begin{figure}
    \centering
    \includegraphics[width=0.45\textwidth]{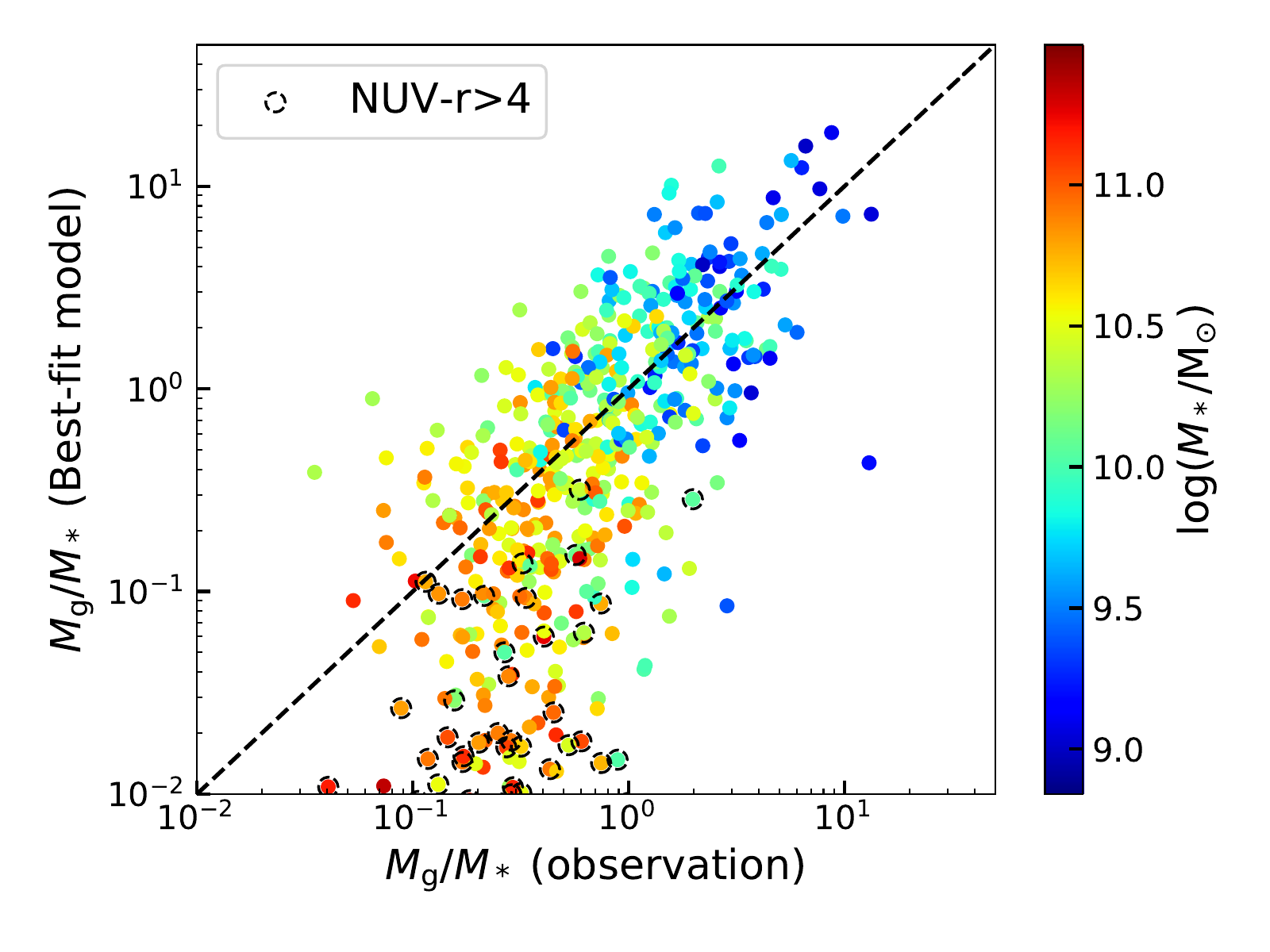}
     \caption{Comparison between the gas to stellar mass ratio predicted by our best-fit model (y axis) and observed in HI-MaNGA (x axis). Dots are our sample galaxies color-coded by their stellar masses. Galaxies with $NUV-r>4$ are identified. The black dash line indicates the 1 to 1 relation, as reference.}
     \label{fig:HI_current}
\end{figure}

\begin{figure}
    \centering
    \includegraphics[width=0.45\textwidth]{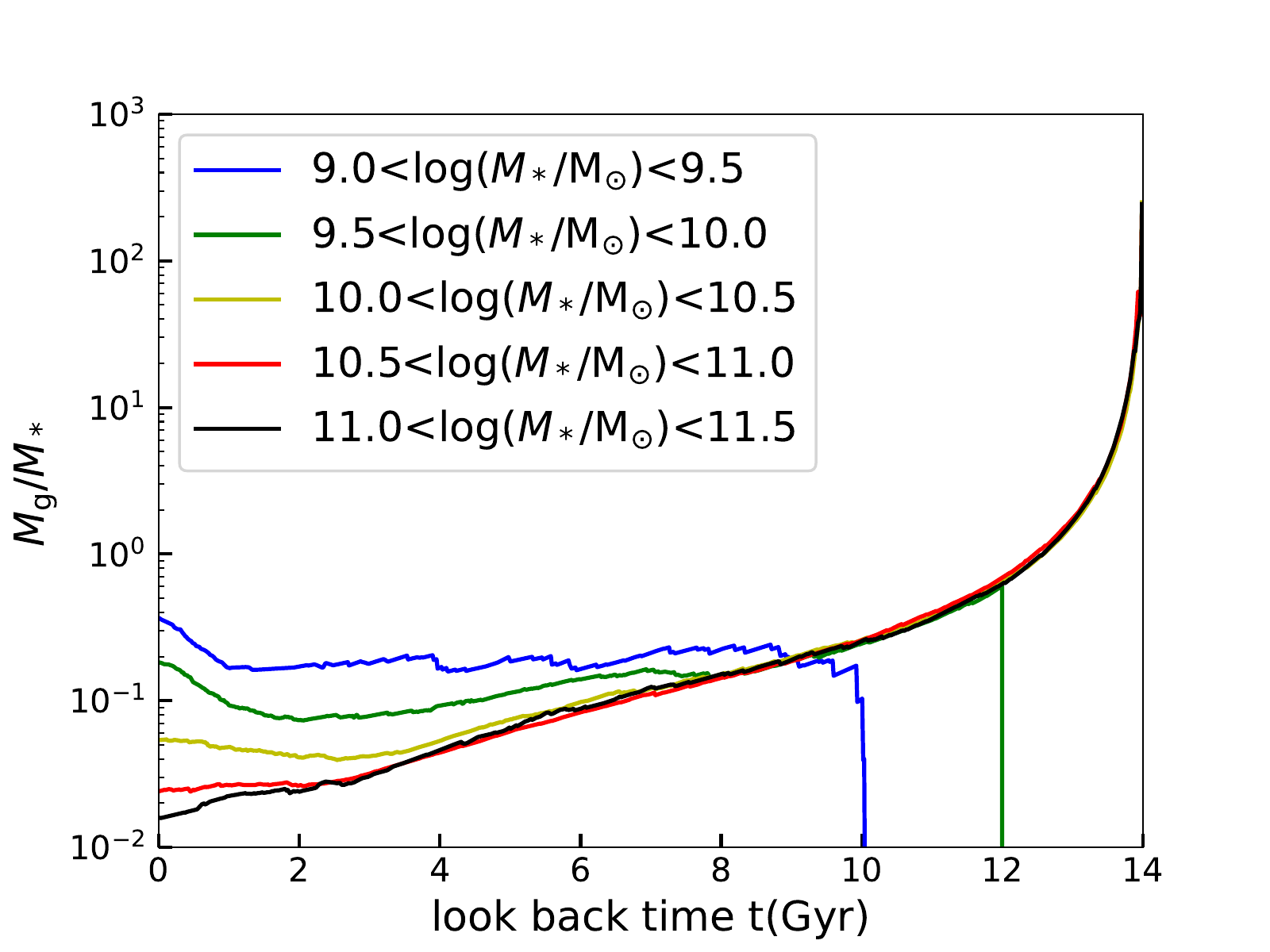}
     \caption{The evolution of gas-to-stellar mass ratios 
     predicted by best-fit models. Lines are medians over the sample galaxies in five stellar mass bins as labeled. The cut-off at early times in low-mass galaxies is an artefact that arises because many of those small galaxies have not started the gas infall process (see \autoref{fig:gasinfall}) and therefore do not have meaningful gas-to-stellar mass ratios.}
     \label{fig:HI_hist}
\end{figure}

In \autoref{fig:HI_current} we compare the observed gas mass to the aperture-corrected predictions of the model, plotted as a fraction of the total stellar mass.  Over two decades in values, there is a good agreement between the model's predictions of present-day gas mass and the independently-measured observed value.  The variation in gas fraction with mass is also readily apparent from this figure, with lower-mass galaxies very much more gas rich. With the help of the model we can now trace this ratio back over time to begin to understand how these differences arose.  Interestingly, as \autoref{fig:HI_hist} shows, $\sim 10\,{\rm Gyr}$ ago all galaxies began with very similar gas fractions. However, the subsequent evolution, with low-mass galaxies' slow-but-steady star formation dependent on a steady supply of gas while high-mass galaxies rapidly depleted theirs, led to the observed differentiation by mass in \autoref{fig:HI_current}.

The tail of galaxies in which the best-fit model under-predicts the observed gas fraction in \autoref{fig:HI_current} warrants a little further investigation.  As the highlighted points in this figure show, many of these objects, which have a reasonably wide range of masses, are extremely red, with $NUV - r > 4$, which places them in the category of "red spiral galaxies" \citep[e.g.][]{Zhou2021}.  It would seem that such systems have largely depleted the gas in their inner regions, so are no longer forming stars.  They still have a reservoir of gas at large radii, which shows up in the HI observations, but the coupling of this material to their inner parts is sufficiently weak that it ends up playing no part in the star-formation history of the galaxy and hence the spectral fitting undertaken in this paper.

As the comparisons in this section have shown, the simple semi-analytic spectral model that we have adopted can reproduce both qualitatively and quantitatively not only results obtained by more ad hoc spectral fitting techniques to the same kind of data, such as the cumulative metallicity distribution function, but also the results derived from totally different types of data, such as direct observations of masses and metallicities of high-redshift galaxies, and radio observations of present-day gas mass. With this reassurance that the model seems to be capturing at least some of the true processes of galaxy formation, we now turn to interpreting the physical meaning of the parameters obtained in the fits. 

\section{Interpretation of the results}
\label{sec:interp}
As described in \autoref{sec:ingredients}, in the simple model that we fit to the data, star formation and chemical evolution in galaxies is regulated by infall and outflow of gas. Since we now have some confidence in the physical significance of the model, we can use the values obtained for the parameters describing these processes to obtain insights into their relative importance in galaxies of different types.  

\subsection{Gas inflow}
\begin{figure}
    \centering
    \includegraphics[width=0.4\textwidth]{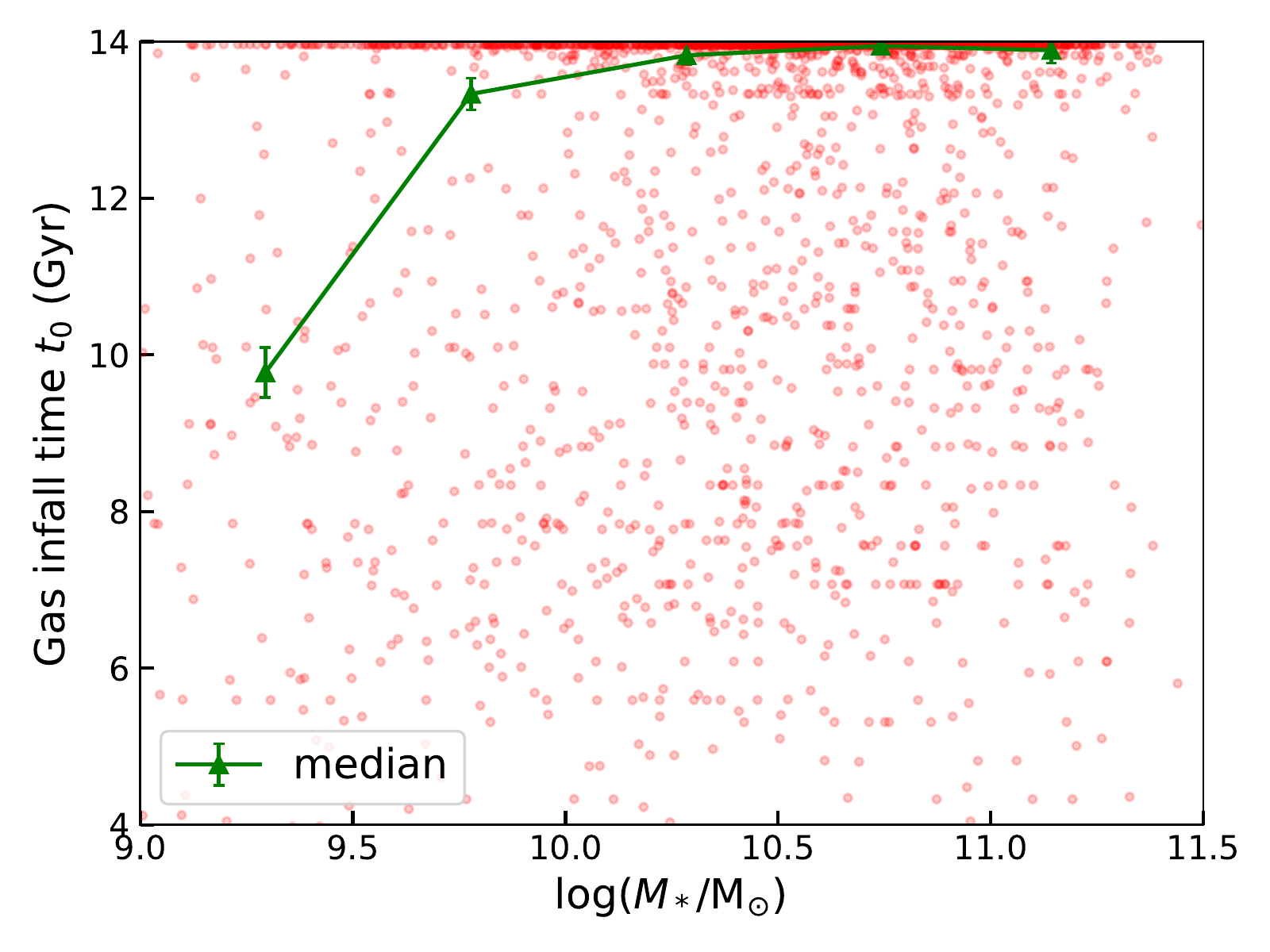}
    \includegraphics[width=0.4\textwidth]{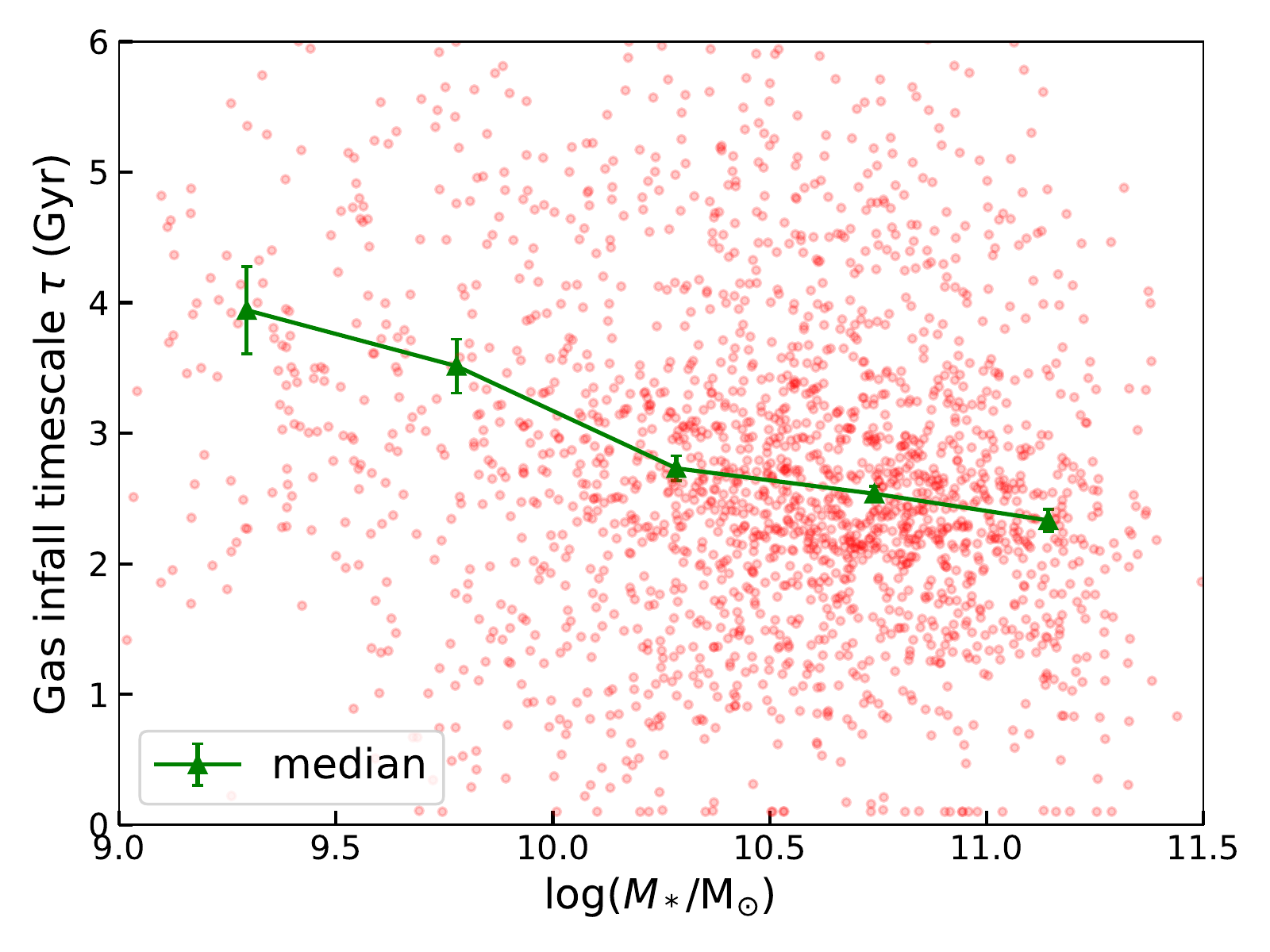}
     \caption{The gas infall start time (top) and timescale (bottom) of our sample galaxies.  Lines are medians over the sample galaxies in five stellar mass bins, with statistical errors shown as error bars.}
     \label{fig:gasinfall}
\end{figure}

In \autoref{fig:gasinfall} we plot the values of the infall parameters that dictate the accretion of gas.  From the top panel it is clear that in the vast majority of galaxies an early start to the accretion process is favoured, with many pushing to the earliest time that our modelling physically allows.  Such saturation is not really a problem, as it is driven simply by the limited ability of spectral data to constrain such old stellar populations, but it does mean that some care is required in interpreting results for individual galaxies.  There is also a clear trend that higher-mass galaxies seem to start this accretion process systematically earlier, which couples to the fact that their star formation begins sooner (see \autoref{fig:sfh}).

The timescale over which this inflow continues, shown in the lower panel of  \autoref{fig:gasinfall}, also depends on galaxy mass, with low-mass systems acquiring gas over a timescale that is almost twice as long, which allows them to enjoy a more extended period of star formation, as seen in  \autoref{fig:sfh}. However, in all cases the infall is far from instantaneous, which is why we find properties like the G-dwarf problem (see \autoref{sec:cmdf}), which show that galaxies across quite a wide range of masses do not behave like simple closed systems.

\begin{figure}
    \centering
    \includegraphics[width=0.4\textwidth]{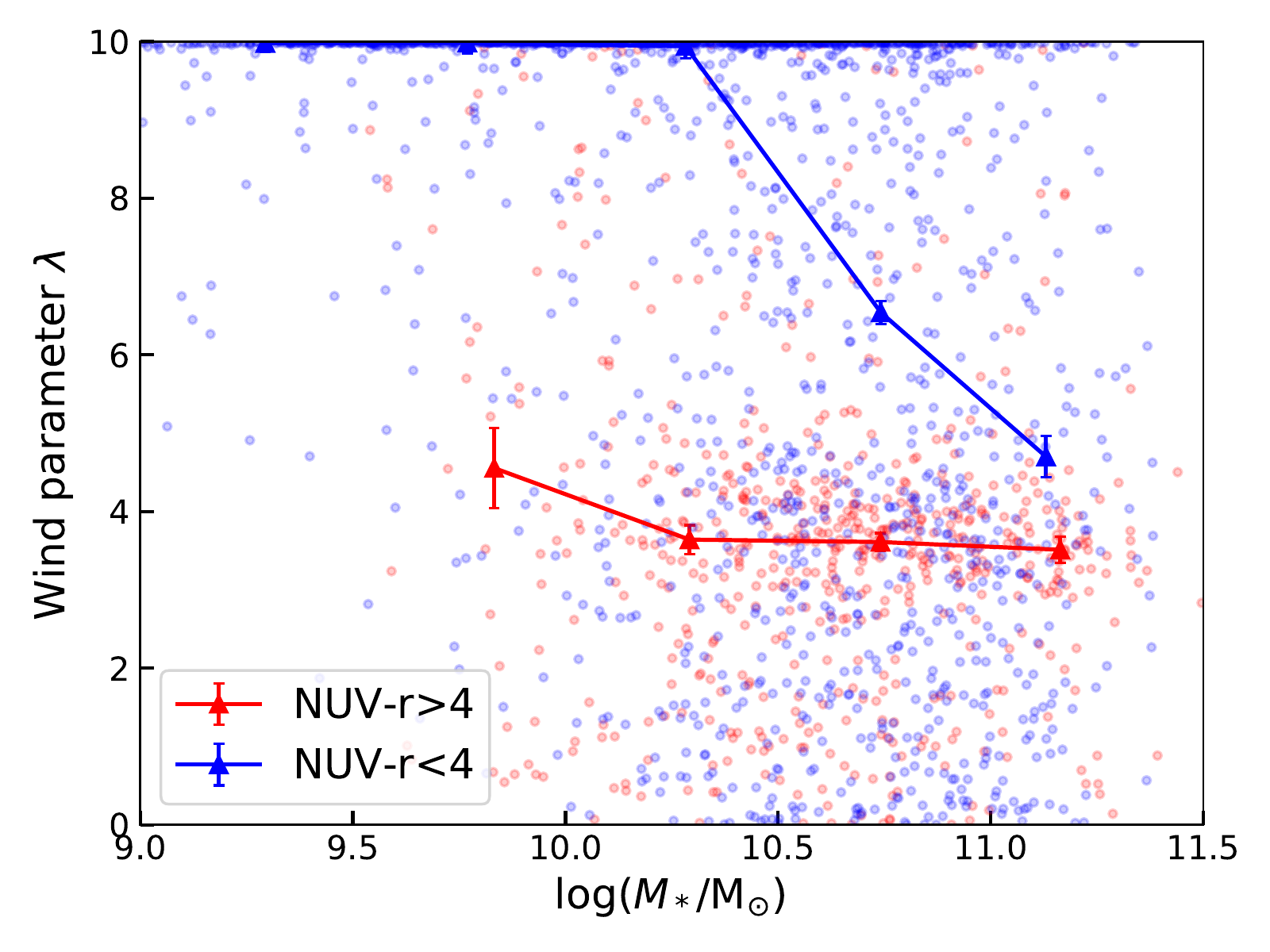}
    \includegraphics[width=0.4\textwidth]{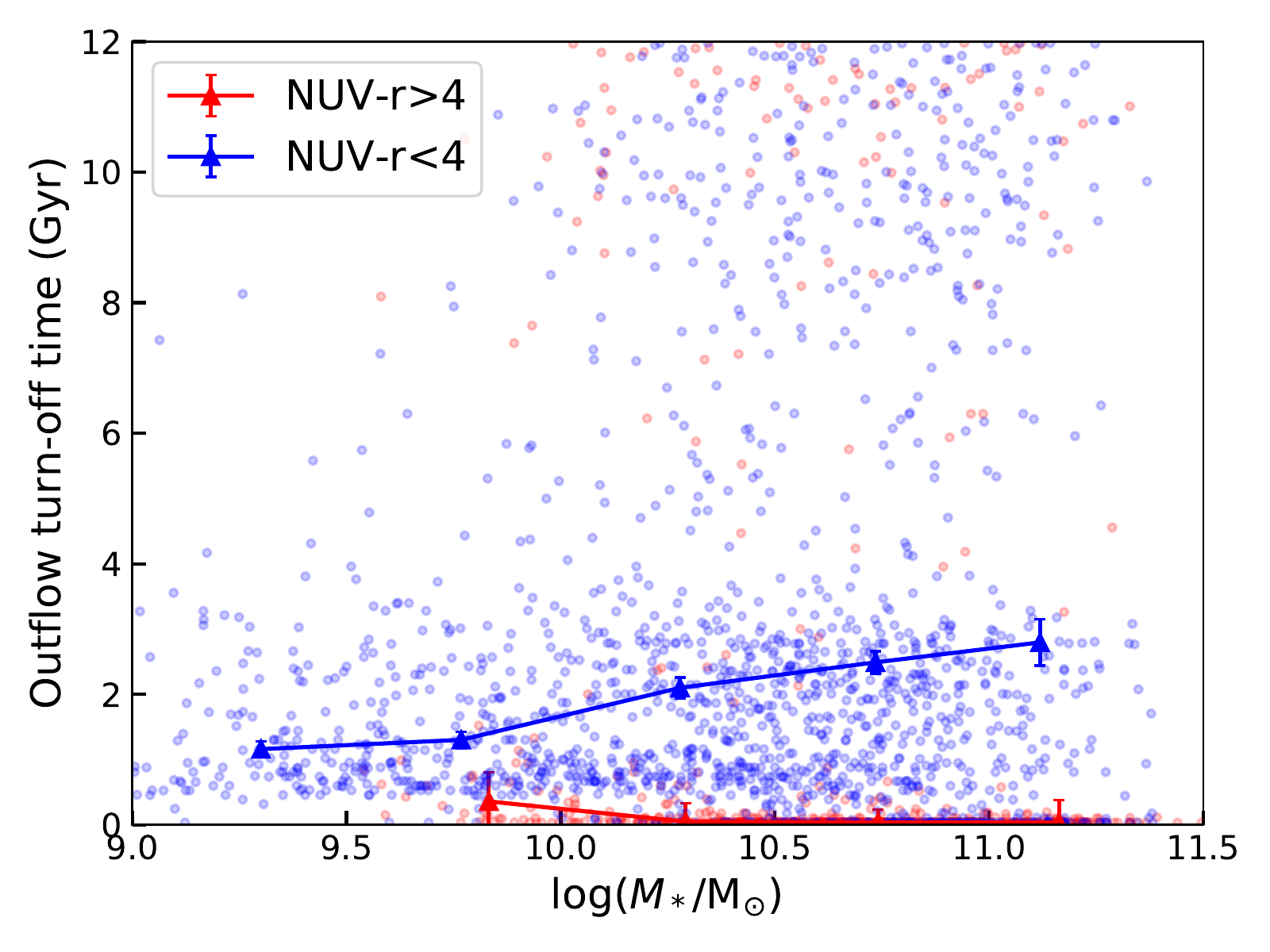}
     \caption{The wind parameter (top) and outflow turn-off time (bottom) of our sample galaxies. Red and blue
     dots are galaxies with $NUV-r>4$ and $NUV-r<4$, with lines showing medians over the sample galaxies in five stellar mass bins.}
     \label{fig:gasoutflow}
\end{figure}

\subsection{Gas outflow}
Outflow is similarly quantified by two parameters in this simplified model, one defining its strength and the other when it switches off.  The value of these parameters vary somewhat with mass, but, as \autoref{fig:gasoutflow} shows, we have found that these variations also depend strongly on colour, with blue spirals ($NUV-r < 4$) showing much stronger variations than red ones ($NUV-r > 4$). For the blue spirals, a wind parameter of $\lambda \sim 10$ for a galaxy with mass around $10^{9.5}{\rm M}_{\odot}$ indicates that only around 10\% of the infalling gas has the chance to turn into stars before being ejected. This strong outflow suppresses the metal enrichment process, keeping their metallicity at a low value as shown in \autoref{fig:ChEH}. The outflow turn-off time shows that their main feedback processes ended in the last $4\,{\rm Gyr}$, which triggers the second major star formation event (see green and blue lines in \autoref{fig:sfh}) and the associated fast metal enrichment (see green and blue lines in \autoref{fig:ChEH}). By contrast, for massive galaxies ($\sim10^{11.0}{\rm M}_{\odot}$), the wind parameter is lower ($\lambda \sim 5$) and even this weaker outflow switches off somewhat sooner.   The weaker and shorter outflow in these more massive galaxies facilitates a quicker metal enrichment. Moreover, from \autoref{fig:model_example_outflow_turnoff}, we see that galaxies with weak outflows would only experience a mild secondary star-formation episode, so that the SFH of massive galaxies only slightly deviates from a monolithic decline (see \autoref{fig:sfh}).

The properties of the red spirals, with $NUV-r>4$, are rather different.  Very few are of low mass, but there is little indication that their parameters vary systematically with mass, with wind parameters of $\lambda \sim 4$ and very low cut-off times which indicate that the wind continues to the present day. Their evolution is thus similar to the model presented in the red line of \autoref{fig:model_example_outflow} -- they quickly formed their stars and enriched in metals on a short timescale, maintaining a constant metallicity ever since. Previous work has shown that the properties of red and blue spirals are different in many respects \citep[e.g.][]{Zhou2021}, and this analysis would suggest that the driving physics may well be differences in the strength and duration of their gas outflows.

We can also compare the values of the outflow parameters to those found by other authors using different constraints.  The range that we find for $\lambda$ are very similar to those obtained by \cite{Spitoni2017}, who used the MZR to constrain the chemical evolution in both star forming and passive galaxies.  The time dependence of the outflows also agrees with \cite{Lian2018mzr} who sought to fit the MZR from both gas and stellar phase simultaneously with their chemical evolution model. Again, the physical plausibility of the parameters that we find, and their agreement with those obtained in other attempts to model the properties of populations of galaxies, indicates that our fits to the life histories of individual galaxies are at least plausible.  

In summary, the complex behavior of gas inflows and outflows regulates both the star formation and metallicity enrichment processes in galaxies. We have sought to paramterize these effects with the simplest model that can capture their essence, and have shown that spiral galaxies formed through varying scenarios for the accreted material. The gas infall in massive galaxies ($\gtrsim 10^{10.5}{\rm M}_{\odot}$) happened very early and with a moderate timescale ($\sim 3$Gyrs). The outflow processes are mild, which induces a gradually decaying SFH and fast metal enrichment in those galaxies. By contrast, the gas infall in less massive galaxies ($\lesssim 10^{10.5}{\rm M}_{\odot}$) occurs later and lasts longer, causing a more extended SFH in them. In addition, outflows in small galaxies are very strong, which suppress early metal enrichment process in these galaxies. After the outflow shuts off $\sim 3\,{\rm Gyr}$ ago, these small galaxies experience a secondary period of star formation, and their gas phase metallicity rapidly increases to the values observed in the local Universe. 

\section{Summary and conclusions}
\label{sec:summary}
In this paper, we have introduced a new method for studying galaxies' chemical and stellar evolution.
We designed a relatively simple model, which encompasses the main process of star formation, gas inflow and outflow, and is suitable for the investigation of systems in which major mergers have not played a significant role. We use both absorption and emission lines to constrain the model, which we fit using the Bayesian inference code {\tt BIGS}.  Extensive testing has confirmed the robustness of the method in both ideal and non-ideal cases, as long as the signal-to-noise ratio of the spectral data is high.

As an initial application of the method, we use it to model a sample of spiral galaxies observed as part of the MaNGA survey, combining the data for each object into a single integrated spectrum to study the global evolution of each system.  The main results of this study are as follows:

\begin{itemize}
\item
The derived SFHs confirm the "down-sizing" formation scenario: massive galaxies ($\sim10^{11}{\rm M}_{\odot}$) 
accumulated 90\% of their masses more than $5\,{\rm Gyr}$ ago, while a significant fraction of the stellar mass in low-mass ($\sim10^{9.5}{\rm M}_{\odot}$) galaxies formed in the last $4\,{\rm Gyr}$.

\item
We find a clear mass dependence in the ChEHs of the galaxies: massive galaxies ($\sim10^{11}{\rm M}_{\odot}$) steadily accumulate their metals since their formation, while the metallicity in low mass galaxies ($\sim10^{9.5}{\rm M}_{\odot}$) evolves slowly over the first $\sim 10\,{\rm Gyr}$, and sharply increases in the last $\sim 4\,{\rm Gyr}$.

\item
The cumulative metallicity distribution function derived from the SFHs and ChEHs also has a clear mass dependence. Only 20\% of the stars in massive ($\sim10^{11}{\rm M}_{\odot}$) galaxies have metallicity lower than Z=0.005, while the fraction increase to around 80\% for low mass galaxies ($\sim10^{9.5}{\rm M}_{\odot}$). This result is consistent with previous studies of the G-dwarf problem, and provides a physically-motivated explanation for it through the derived combination of gas inflow and outflow.

\item
The mass--metallicity relation is found to evolve only slowly down to a redshift of $z \sim 0.3$, and metallicity at any given mass then increases rapidly more recently.  The relation is also seen to flatten at the high-mass end as metallicity saturates in these galaxies.  These results, driven by the physics of the model applied to nearby galaxies, are found to agree with those obtained from direct observations of high-redshift samples. 

\item
The model also successfully predicts the observed gas-to-stellar mass ratios for the galaxies in the sample. This ratio shows a clear mass dependence, such that galaxies with low stellar masses have higher gas mass fractions. We can trace this ratio back in time with the model, and find that all galaxies began with similar ratios $\sim 10\,{\rm Gyr}$ ago, but that their different physical drivers of star formation and gas flow led to the variations in this ratio that we see today.

\item
The evolution described above is strongly regulated by the derived gas inflow and outflow processes. High-mass galaxies have an earlier gas infall time and shorter infall timescale, which dictates the earlier star formation and metallicity enrichment processes seen in these galaxies.
In addition, the outflows in massive galaxies are weaker than in less massive ones, and in many massive galaxies the outflow turns off early in their lifetimes. The evolution of massive galaxies is thus less affected by the gas outflow. By contrast, strong outflows suppress the early metallicity enrichment process in low-mass galaxies.
Starting from around $4\,{\rm Gyr}$ ago and especially in the recent 2 Gyr, this strong outflow turned off in low-mass systems, triggering a secondary phase of star-formation and relatively rapid metallicity enrichment. 

\end{itemize}

The success of this initial application of semi-analytic spectral fitting gives us some confidence in its credibility.  The fact that it reproduces a range of entirely independent results indicates that it is offering much more than just a convenient spectral fitting tool, but is actually capturing much of the physics that drives star formation and chemical evolution.  For this initial proof of concept, we have designed a simple physically-motivated chemical evolution model, and have used as much information as possible to constrain it. In the future, we plan to continue refining the model. For example, in this work we have focused on star-forming, late-type galaxies that have at least some level of ionised gas to provide necessary additional constraints on the fits, but it may be possible to use quite general scaling relations as auxiliary constraints to gas phase metallicities and star formation rates, so that the method can also be applied to galaxies such as S0s that are not currently star forming. In addition, in this work we have assumed a merge-free evolutionary history, which would probably not be valid when studying elliptical galaxies, but there is nothing fundamental to the approach that prevents the adoption of a model that includes mergers. Moreover, although we have focused on the integral properties of galaxies in this paper, a similar approach could be applied to spatially-resolved spectra of adequate quality, going beyond the global properties of galaxies to investigate how they vary internally. In such a case, we would have to consider the potential interplay between the different parts of the galaxy -- in the case of the Milky Way, for example, it is known that large-scale metal-enriched gas flows are necessary to explain the evolution of its stellar population \citep[e.g.][]{Spitoni2011,Vincenzo2020}. This complication would require the introduction of further parameters describing the coupling between regions, but with the potential reward of shedding new light on galaxies' internal evolutionary processes.  The novel tool of semi-analytic spectral fitting offers an exciting bridge between theoretical evolutionary models and the observational properties of galaxies, and we look forward to its on-going refinement and exploitation beyond this initial study.

\section*{Acknowledgements}
SZ, MRM and AAS acknowledge financial support from the UK Science and Technology Facilities Council (STFC; grant ref: ST/T000171/1).

For the purpose of open access, the authors have applied a creative commons attribution (CC BY) to any journal-accepted manuscript.

Funding for the Sloan Digital Sky Survey IV has been provided by the Alfred P. 
Sloan Foundation, the U.S. Department of Energy Office of Science, and the Participating Institutions. 
SDSS-IV acknowledges support and resources from the Center for High-Performance Computing at 
the University of Utah. The SDSS web site is www.sdss.org.

SDSS-IV is managed by the Astrophysical Research Consortium for the Participating Institutions of the SDSS Collaboration including the Brazilian Participation Group, the Carnegie Institution for Science, Carnegie Mellon University, the Chilean Participation Group, the French Participation Group, Harvard-Smithsonian Center for Astrophysics, Instituto de Astrof\'isica de Canarias, The Johns Hopkins University, Kavli Institute for the Physics and Mathematics of the Universe (IPMU) / University of Tokyo, Lawrence Berkeley National Laboratory, Leibniz Institut f\"ur Astrophysik Potsdam (AIP), Max-Planck-Institut f\"ur Astronomie (MPIA Heidelberg), Max-Planck-Institut f\"ur Astrophysik (MPA Garching), Max-Planck-Institut f\"ur Extraterrestrische Physik (MPE), National Astronomical Observatories of China, New Mexico State University, New York University, University of Notre Dame, Observat\'ario Nacional / MCTI, The Ohio State University, Pennsylvania State University, Shanghai Astronomical Observatory, United Kingdom Participation Group, Universidad Nacional Aut\'onoma de M\'exico, University of Arizona, University of Colorado Boulder, University of Oxford, University of Portsmouth, University of Utah, University of Virginia, University of Washington, University of Wisconsin, Vanderbilt University, and Yale University.

\section*{Data availability}
The data underlying this article were accessed from: SDSS DR17 \url{https://www.sdss.org/dr17/manga/}. The derived data generated in this research will be shared on request to the corresponding author.

\bibliographystyle{mnras}
\bibliography{szhou} 

\bsp	
\label{lastpage}
\end{document}